\documentclass[11pt]{article}
\usepackage{amsmath, amssymb, amscd, amsthm, amsfonts}
\usepackage{graphicx}
\usepackage{hyperref}
\usepackage{verbatim}
\usepackage{float}
\usepackage{booktabs}
\usepackage{algorithm}
\usepackage{algorithmic}
\title{Security Analysis of Pairing-based Cryptography}
\author{Xiaofeng Wang\footnote{College of Computer, National University of Defense Technology, Changsha 410000, China}, Peng Zheng\footnote{Corresponding author, zhengpeng21a@nudt.edu.cn}, Qianqian Xing\footnote{Corresponding author,  xingqianqian12@nudt.edu.cn} 
}

\date{2023.06}

\begin{document}

\maketitle

\begin{abstract}
Recent progress in number field sieve (NFS) has shaken the security of Pairing-based Cryptography.  For the discrete logarithm problem (DLP) in finite field, we present the first systematic review of the NFS algorithms from three perspectives: the degree $\alpha$, constant $c$, and hidden constant $o(1)$ in the asymptotic complexity $L_Q\left(\alpha,c\right)$ and indicate that further research is required to optimize the hidden constant. Using the special extended tower NFS algorithm, we conduct a thorough security evaluation for all the existing standardized PF curves as well as several commonly utilized curves, which reveals that the BN256 curves recommended by the SM9 and the previous ISO/IEC standard exhibit only 99.92 bits of security, significantly lower than the intended 128-bit level. In addition, we comprehensively analyze the security and efficiency of BN, BLS, and KSS curves for different security levels. Our analysis suggests that the BN curve exhibits superior efficiency for security strength below approximately 105 bit. For a 128-bit security level, BLS12 and BLS24 curves are the optimal choices, while the BLS24 curve offers the best efficiency for security levels of 160bit, 192bit, and 256bit.

\end{abstract}
\section{Introduction}\label{section-introduction}
Since Boneh and Franklin proposed identity-based encryption in 2001, the research on pairings has gained momentum. Pairing-based cryptography(PBC) and pairing-friendly(PF) curves have been extensively studied and have led to the development of several unique protocols that cannot be achieved by traditional methods. These include digital signatures\cite{2002An,2003ID}, key exchange protocols\cite{2007Identity}, chameleon hashing\cite{2009Identity}, short/aggregate/verifiably encrypted/group/ring/blind signatures\cite{2001Short,2003Aggregate,2006New,2002ID,2004An}, and broadcast encryption\cite{18}. Several candidates of PF curves and PBC have been recommended by various standards including ISO/IEC 11770-3, 14888-3, 18033-5, IEEE P1363.3, IETF RFC 6508 and Chinese national cryptography standard SM9. Moreover, the PBC are widely used by the Trusted Computing Group (TCG), FIDO Alliance, W3C, Intel SGX, and blockchain projects such as Ethereum, Algorand, China Network, and DFINITY.
	
	However, the security of the pairings have not received sufficient attention, leading to incorrect usage. Recent remarkable progress in solving the discrete logarithm problems(DLP) over finite fields has shaken the security of PF curves. Hence, it is imperative to overview the progress of the finite field security analysis and conduct a thorough reassessment of the security of the PF curves. Furthermore, we need to update the current standards with appropriate PF curves based on the latest evaluation results.
	
	A Cryptography pairing is a non-degenerate bilinear map $e : \mathbb{G}_1 \times \mathbb{G}_2 \rightarrow \mathbb{G}_T$. $\mathbb{G}_1$ and $\mathbb{G}_2$ are order-r additive groups of an elliptic curve defined over a finite field $F_p$ or one of its extensions, $\mathbb{G}_T$ is the order-r multiplicative cyclic group defined on the extended field $F_{p^k}^\ast$ where $k$ is called the embedding degree and p is called characteristic. The security of bilinear pairing in cryptosystems relies on the hardness of solving the DLP on both the elliptic curve group (curve-side security) and the finite field extension multiplication group (field-side security). The curve-side security can be addressed by the Pollard's rho algorithm, while the field-side security remains a challenging problem that has attracted significant attention in the academic community.
	
	The most advanced algorithm for solving the DLP over finite fields with non-small characteristic is the number field sieve (NFS) algorithm. In recent years, the NFS algorithm has made significant progress in solving the DLP in the extended fields $F_{p^k}$, posing a serious threat to the security of PBC. However,  the existing survey\cite{2017Challenges} of  NFS algorithms can not cover the latest progress of optimizing the hidden constant for a more precise security assessment, which leads to misuse of various NFS algorithms. Under the latest SexTNFS\cite{10.1007/s00145-018-9280-5} algorithm, the current international standard curves do not give an accurate assessment of the security, only gives an estimate, which undermines the confidence of PBC users in the security. In addition, to select PF curves with security levels of 128, 160, 192, and 256 bits, researchers and some standards follow trends without considering the trade-off between security and efficiency, which leads to improper curve selection. 
	
	This paper presents a systematic review of the state-of-the-art in finite field security analysis for Pairing-based Cryptography and provides a comprehensive evaluation of the current PF curves standards and applications. In addition, we give an assessment of the security and efficiency of various PF curves at different security levels. The main contributions of this paper are as follows:

	1. For field-side security analysis, we give the first systematic review of the  NFS algorithms from three perspectives: the degree $\alpha$, the constant $c$, and the hidden constant o(1) in the time complexity $L_Q\left(\alpha,c\right)$. Our research indicates that the complexity of DLP in finite field can be reduced from $L_{p^n}\left(1/2\right)$ to $L_{p^n}\left(1/3,1.526\right)$, while the special extended tower NFS algorithms can give the minimal constant c with the most optimal polynomial. However, the hidden constant can still be optimized, which needs further research.
	
	2.Using the special extended tower NFS algorithm, we conduct the first thorough security evaluation for all the existing standardized PF curves as well as several commonly utilized curves. Our analysis reveals that the BN256 curves of SM9 and ISO/IEC exhibit a security strength of only 99.92 bits, significantly lower than the intended 128-bit level. However, the newly proposed IEC/ISO curves are capable of supporting security levels of either 128 bits or 192 bits. 
	
	3. We comprehensively analyze the security and efficiency of BN, BLS, and KSS curves in different security levels. Considering the balance between security on the curve side and field side, BLS12, KSS16, KSS18, and BLS24 achieve respectively optimal security efficiency at 121bit, 141bit, 164bit, and 183bit security strength. The BN curve exhibits superior efficiency for security strength below approximately 105 bit. For a 128-bit security level, BLS12 and BLS24 curves are the optimal choices, while the BLS24 curve offers the best efficiency for security levels of 160bit, 192bit, and 256bit. Since the ISO/IEC's BN512 curve is relatively less of efficiency with a security strength of 138 bits, we recommend to use the BLS24 curve for the same security strength.

	This paper is structured as follows: in §2, an overview of the NFS is provided, with an analysis of the main factors that impact its security. In §3, we present a review of the state-of-the-art in finite field security analysis from three perspectives, and a comparison of the evolution of NFS is presented. §4 provides an introduction to the popular BN, BLS, and KSS curves, including their key parameters. In §5, the security of international standards for PF curves is analyzed, along with the efficiency of each curve under specific security levels. Finally, in §6, concluding remarks are made.

 	\section{Overview of Number Field Sieve}
	This chapter provides an overview of the NFS algorithm and examines the key factors that impact its security. The DLP and the decomposition of large numbers over finite fields are intertwined. In 1988, Pollard\cite{10.1007/BFb0091536} introduced NFS as a solution for decomposing large numbers. Gordon\cite{Gordon1993Discrete} later proposed applying NFS to solve the DLP.
 
 NFS is an index calculus algorithm for computing DLP in the finite field $F_Q$, where $Q = p^n$ for some prime $p$. NFS is inspired by the quadratic sieve algorithm and utilizes the concepts of factor basis and smoothness. Given $g,s\in F_Q$ where $g^x=s$ for some integer $x$, the discrete logarithm of $s$ can be found by selecting a random integer $e$ from $[0,Q-1]$, computing $z=s^e$, and finding a factor basis in which $z$ is smooth. The discrete logarithm of $s$ can then be calculated as $\log_g{\left(s\right)}=\log_g{\left(z\right)/e}$. In this way, the problem of finding the discrete logarithm of $s$ is transformed into the problem of finding the discrete logarithm of $z$, where $z=p_1^{e_1}\times\cdots\times p_k^{e_k}$, there is linear relationship $\log_g{\left(z\right)}=e_1\log_g{p_1}+\cdots+e_k\log_g{p_k}$. If enough of these relationships are obtained, we can solve for the discrete logarithm of $z$ by solving the linear equations. NFS consists of four main phases: polynomial selection, relation collection (sieving), linear algebra, and individual discrete logarithm calculation.
	
	Polynomial selection is a critical step in NFS, as the quality of the polynomial directly affects the number of relations obtained in the subsequent phases. The appropriate factor basis and sieving polynomial are selected based on specific parameters to obtain enough relations through the lattice sieve\cite{10.1007/BFb0091538}. In the linear algebra phase, the discrete logarithm of the factor basis elements can be obtained by solving the equations obtained from the relation collection phase. Finally, based on the discrete logarithm of the factor basis elements, individual discrete logarithms can be calculated.

\begin{figure}[htbp]
		\centering		\includegraphics[width=0.5\textwidth]{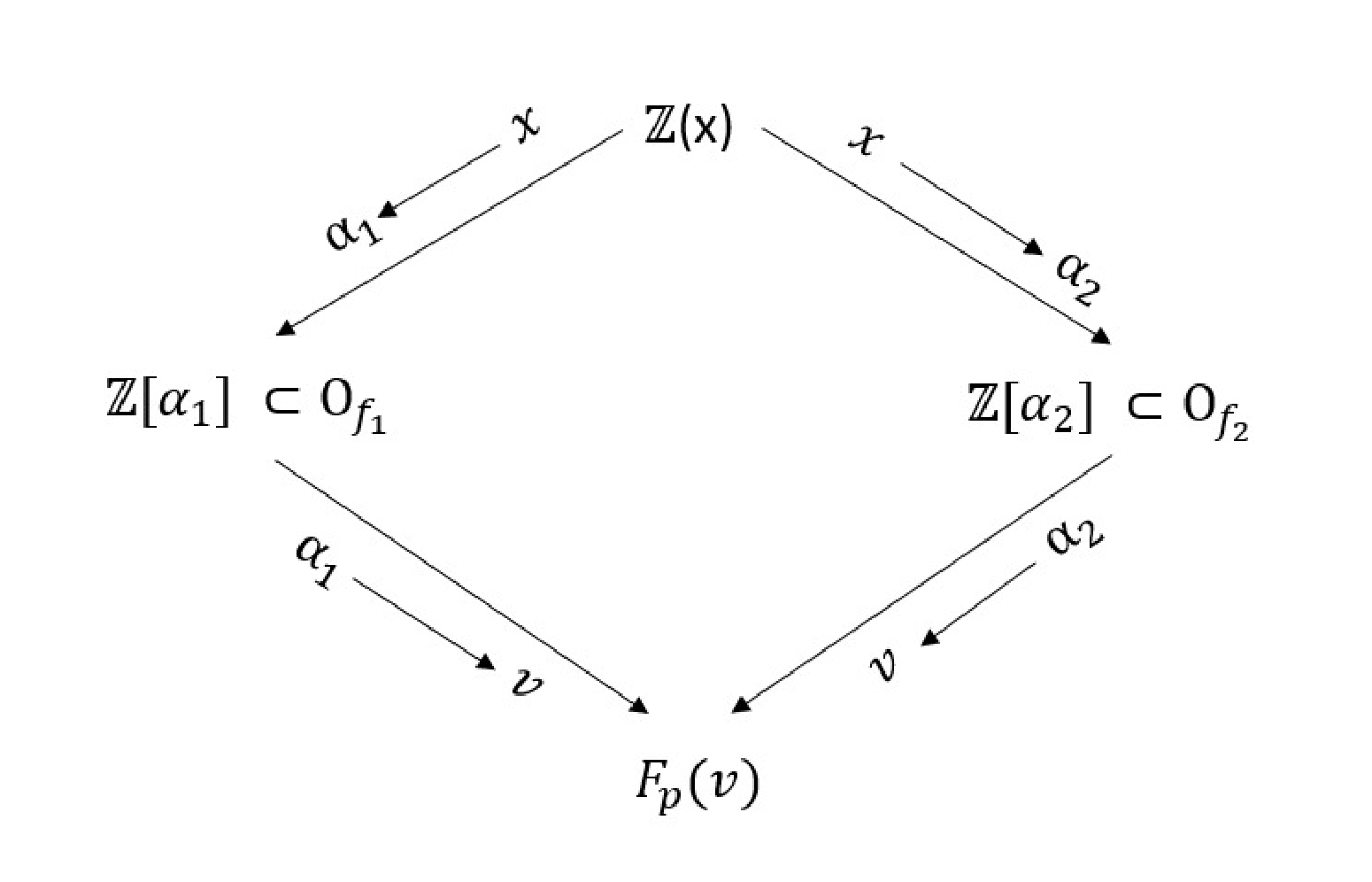}
		\caption{The basic principle of NFS.}
		\label{Figure1}
	\end{figure}

	NFS aims to establish smooth relations between number fields by utilizing the residual field isomorphism of first-order prime ideals. To achieve this, we select irreducible polynomials $f_1(x)$ and $f_2(x)$ on the integer ring, with roots $\alpha_1$ and $\alpha_2$ in $\mathbb{C}$, respectively. We require that $f_1(x)$ and $f_2(x)$ have a common irreducible factor $h(x)$ of degree $n$ over $F_p$ when considered modulo $p$. The number fields $K_{f_1}=Q([x]/(f_1))$ and $K_{f_2}=Q([x]/(f_2))$ are constructed using $\alpha_1$ and $\alpha_2$, respectively. The field $F_{p^n}$ is represented by $h(x)$, and $v\in F_{p^n}$ denotes the roots of $h(x)$. $O_{f_1}$ and $O_{f_2}$ are the rings of integers over $K_{f_1}$ and $K_{f_2}$, respectively. The rationale of the NFS is illustrated in the commutative diagram (Figure \ref{Figure1}), which depicts two homomorphisms $K_{f_1}\rightarrow F_{p^n}$ and $K_{f_2}\rightarrow F_{p^n}$ given by $\alpha_1\rightarrow v$ and $\alpha_2\rightarrow v$, respectively.

	NFS works on the ring of integers $O_{f_1}$ and $O_{f_2}$ instead of $K_{f_1}$ and $K_{f_2}$. The factor basis is taken to be some prime ideal of $O_{f_1}$ and $O_{f_2}$ with small norm. The size of the factor base is chosen to balance the times for relation collection and linear algebra. To obtain enough relations in a given range, sieving polynomials $\varphi(x)$ are considered. The principal ideals $\varphi\left(\alpha\right)O_{f_1}$ and $\varphi\left(\beta\right)O_{f_2}$ are considered B-smooth if they simultaneously decompose into products of prime ideals in the factor basis. This occurs when the norms $Res(f_1,\varphi)$ and $Res(f_2,\varphi)$ are both B-smooth. The Schirokauer map\cite{2005Virtual} is used to convert each relation into a linear equation consisting of the virtual logarithms of the factor basis ideals. Once enough relational equations are obtained, the discrete logarithm of the factor basis elements can be obtained by solving the equations. Finally, the individual discrete logarithm is solved through the discrete logarithm of the factor basis.
	
	The complexity of the NFS algorithm is mainly determined by the relation collection and computing linear algebra stages. The product of the norms, $\left|Res(f_1,\varphi)\right|\times\ \left|Res(f_2,\varphi)\right|$, significantly impacts the probability of finding a relation and the cost of this step. The polynomials $f_1$ and $f_2$ are key factors affecting the norm, as it depends on their degree and coefficients. While smaller degrees and coefficients can ensure a smaller norm, current polynomial selection methods involve trade-offs between degree and coefficients. Barbulescu and Pierrot\cite{barbulescu_pierrot_2014} have provided an analytical explanation for the negligible time required to compute an individual discrete logarithm.

 	\section{Evolution of Number Field Sieve Algorithms}
	In this section, we provide an overview of the development of the Number Field Sieve (NFS) algorithm and examine the differences and connections between its various variants.  Let characteristic $p$ be expressed as $p=L_Q\left(\alpha,c\right)$, where $\alpha,c>0$. Based on the value of $\alpha$, finite fields $F_Q$ can be categorized into four types: small characteristic for $\alpha\le 1/3$, medium characteristic for $1/3<\alpha<2/3$, boundary for $\alpha=2/3$, and large characteristic for $\alpha>2/3$. PBC always uses medium characteristic finite fields. Hence, we primarily focus on NFS in non-small characteristic finite fields whose asymptotic complexity can be expressed as:
	
	\centerline{$L_Q\left(\alpha,c\right)=\exp{\left(\left(c+o\left(1\right)\right)\left(logQ\right)^\alpha\left(loglogQ\right)^{1-\alpha}\right)}$}
	
\begin{figure*}[htbp]
		\centering
		\includegraphics[width=\textwidth]{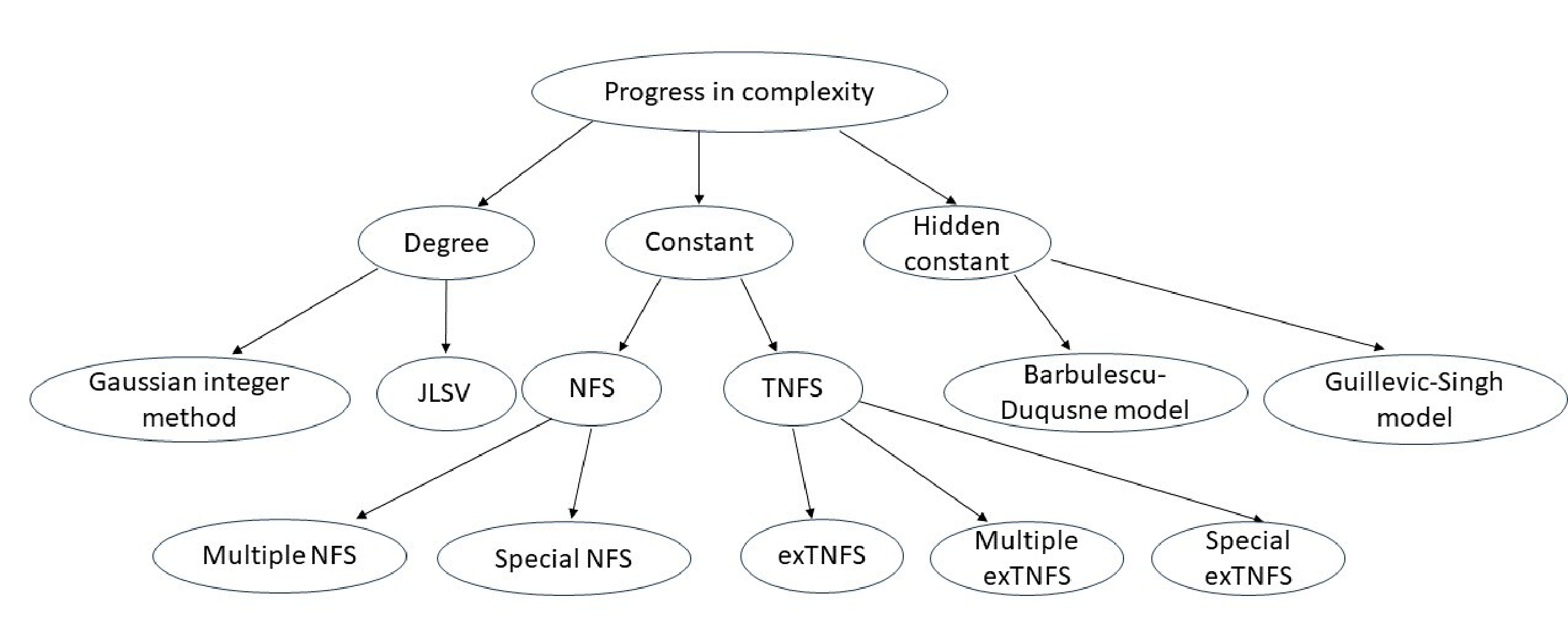}
		\caption{Development of NFS.}
		\label{figure2}
	\end{figure*}

	The initial optimization of NFS mainly focused on the degree $\alpha$ , while subsequent research efforts were directed towards optimizing the constant $c$. Notably,  The tower NFS(TNFS) has made significant progress in this regard. The latest research has focused on optimizing the hidden constant $o(1)$. As shown in Figure \ref{figure2}, we investigate and survey the evolution of NFS from three aspects: degree $\alpha$, constant $c$, and hidden constant $o(1)$. The following three subsections provide a detailed discussion of these aspects.

	\subsection{Improvement of the degree $\alpha$}
	
	The sub-exponential time algorithm for discrete logarithm originated from the Gauss integer method, which was first proposed in 1986 and later realized by Coppersmith \cite{Don1986Discrete}. This algorithm represents the first breakthrough in discrete logarithm calculation methods, with a time complexity of $L_Q\left(1/2,1\right)$. Its advantage lies in its good implementation performance, which has dominated implementations for a long time. In 1993, Gordon\cite{1993Discrete} proposed the NFS method for discrete logarithm, which made a significant breakthrough in solving discrete logarithms over $F_p$. However, the algorithm is difficult to implement. Schirokauer improved Gordon's method by introducing the $\lambda$ mapping \cite{a2ab88eb-7078-31e8-9d80-1ed5ab69777e}, which transformed the relation equation between prime ideals into the relation equation between elements and solved the problem of unit calculation. This algorithm has both theoretical significance and ease of implementation.
	
	In 2000, Schirokauer\cite{2000Using} proposed the use of the number field method for computing the DLP on the extended field. However, the Schirokauer NFS method requires solving a lot of equations, which is a major bottleneck in computing discrete logarithms. In 2003, Lercier and Joux\cite{2003Improvements} improved the Schirokauer algorithm in two ways. Firstly, they improved the selection of polynomials by generating a polynomial of order $d+1$ and using lattice reduction to generate the coefficients of the second polynomial of order $d$. This is a generalization of the Gauss integer method. Secondly, for a fixed module $p$, only one equation needs to be solved, instead of solving a very large linear system for each logarithm. This reduces the time required to solve a individual discrete logarithm.
	
	In 2006, Joux et al.\cite{2006The} enhanced Joux and Lercier's function field sieve and improved the construction of number field in NFS, along with the selection of sieve space. They proposed a polynomial selection method, JLSV1, which is particularly suitable for the medium characteristic case. This method generates two polynomials, $f_1$ and $f_2$, of the same degree n, with coefficients of size $O(\sqrt p)$ each. The computational complexity of discrete logarithm in finite field is heuristically reduced from $L_{p^n}\left(1/2\right)$\cite{1993A} to $L_{p^n}\left(1/3\right)$.
	\subsection{Improvement of the constant c}
	Polynomial selection is a crucial step in the NFS algorithm, and its quality significantly impacts the value of the constant c in $L_{p^n}\left(\alpha,c\right)$. Recent advancements in polynomial selection have resulted in a notable reduction in the value of c. In this context, we examine the impact of various polynomial selection methods on the constant c in complexity.
	\subsubsection{NFS}
	In 2015, Razvan Barbulescu et al.\cite{2015Improving} proposed two novel methods for selecting polynomials in the context of non-prime finite field discrete logarithms. These methods, namely the Generalized Joux-Lercier (GJL) and Conjugation, optimize the coefficients and degree of the polynomials, resulting in lower upper bounds on the norm. The Conjugation method is competitive when the prime characteristic $p$ is smaller (medium prime case), while the GJL method is better suited for larger prime characteristics.bThe GJL method aims to minimize the degrees of both $f_1$ and $f_2$ to n+1 and n, respectively, but it fails to achieve small coefficients of $f_2$. On the other hand, the Conjugation method sets $f_1$ with a higher degree of 2n instead of n+1, enabling the coefficients of $f_1$ to have size $O(1)$ and constraining the coefficients of $f_2$ to be bounded by $O(\sqrt p)$. This method achieves a complexity of $L_Q\left(1/3,\sqrt[3]{96/9}\right)$ in the case of medium prime characteristics.
	
	Barbulescu and Pierrot\cite{barbulescu_pierrot_2014} introduced the concept of using multiple polynomials $f_i$, i = 1,2,...,D in the NFS algorithm, which is known as the Multiple Number Field Sieve (MNFS). By replacing the two polynomials $f_1$ and $f_2$ in NFS with $D$ number fields and increasing the cardinality of elements in the smoothness base, the probability of an element being doubly smooth increases by a factor of $D^2$. Pierrot\cite{C2015The} adapted the MNFS to the Conjugation algorithm. These two methods lead to a decrease in complexity. Sarkar and Singh proposed a new polynomial selection method called Algorithm-$\mathcal{A}$\cite{PalashSarkar2016New}, which improves upon the GJL and Conjugation methods. This method has two key parameters, $d$ and $r$, where $d$ is a divisor of the extension degree $n$ and $r \geq n/d$. When $d=1$, Algorithm-$\mathcal{A}$ is equivalent to the GJL method, and when $d=n$ and $r=k=1$, it is equivalent to the Conjugation method. This method provides a trade-off between the above two methods. It achieves a complexity of $L_Q\left(1/3,{(8(9+4\sqrt6)/15)}^{1/3}\right)$ in the case of medium prime characteristics.
	
	PBC systems often use a polynomial representation $P(u)$ for the characteristic $p$, where the degree $\tau$ of $P(u)$ is small, $u$ is sparse, and the coefficients are small. For this specific case, Joux and Pierrot\cite{2014The} proposed a modification to the polynomial selection stage in NFS and introduced the SNFS algorithm. Firstly, they selected an irreducible polynomial $f_1$ of degree $n$ in $F_p$ as $f_1=x^n+W(x)-u$, where $W(x)$ has a small degree and coefficients of 0, -1, or 1. The coefficient of $f_1$ is bounded by $p^{1/\tau}$. Then, they chose $f_2$ as $f_2=P(x^n+W(x))$, with $f_2(x)=P(f_1(x)+u)\equiv P(u)=p(mod\ f_1(x))$. The degree of $f_2$ is $\tau n$, and the bound on the coefficient is $O({(log\ n)}^\tau$. The asymptotic complexities with medium case is $L_Q(1/3,(64/9)·(\tau+1)/\tau^{1/3})$.

	\subsubsection{TNFS}
	The Tower Number Field Sieve (TNFS) algorithm was first initially by Schirokauer\cite{2000Using} in 2000. In 2015, Barbulescu et al.\cite{2015The} modified it and pointed out that TNFS can further reduce the security of PBC. Kim and Barbulescu proposed a new variant exTNFS\cite{2016Extended} based on TNFS, which has a profound impact on the medium prime case. Unlike the classical NFS algorithm involves two polynomials $f_1$, $f_2$ defined on $\mathbb Z[x]$ and have a common degree $n$ irreducible factor modulo $p$,  TNFS algorithm employs two polynomials $f_1$, $f_2$ defined on a ring of $R$ which is of the form $R = Z[t]/h(t)$, $h(t)$ is a monic irreducible  polynomial of degree n. $f_1$ and $f_2$ have a common root modulo $\mathfrak{p}$ in $R$, where $\mathfrak{p}$ is a unique ideal above $p$ in $R$. $F_{q^n}$ is obtained as $R/\mathfrak{p}$, and there exist two ring homomorphisms from $R[x]/f_1(x)$ and $R[x]/f_2(x)$ to $F_{q^n}$.
	
	When the extension degree $n$ is composite and $n=\eta\kappa$, $gcd(\eta,\kappa)=1$, the security analysis of exTNFS over finite fields with medium characteristics has an important impact. The bacic ideal is $F_{p^n}=F_{(p^\eta)^\kappa}$. To construct such a field, a monic irreducible polynomial $h(t)$ of degree $\eta$ over $F_p$ is first chosen. Then, the quotient ring $R:=\mathbb{Z}[t] / h(t)$ and the field $F_{p^\eta}=R/pR$ are defined. Two irreducible polynomials $f_1$ and $f_2$ over $R$ and $F_{p^\eta}$ are selected, such that their modulo $p$ share a common irreducible factor $\phi(x)$ of degree $\kappa$ over $F_{p^\eta}$. there are two ring homomorphisms from $R[x]/f_1(x)$ and $R[x]/f_2(x)$ to $F_{p^n}=F_{p^{\eta \kappa}}=(R / p R) / \phi(x)$.

	\begin{figure}[htbp]
		\centering
		\includegraphics[width=0.35\textwidth]{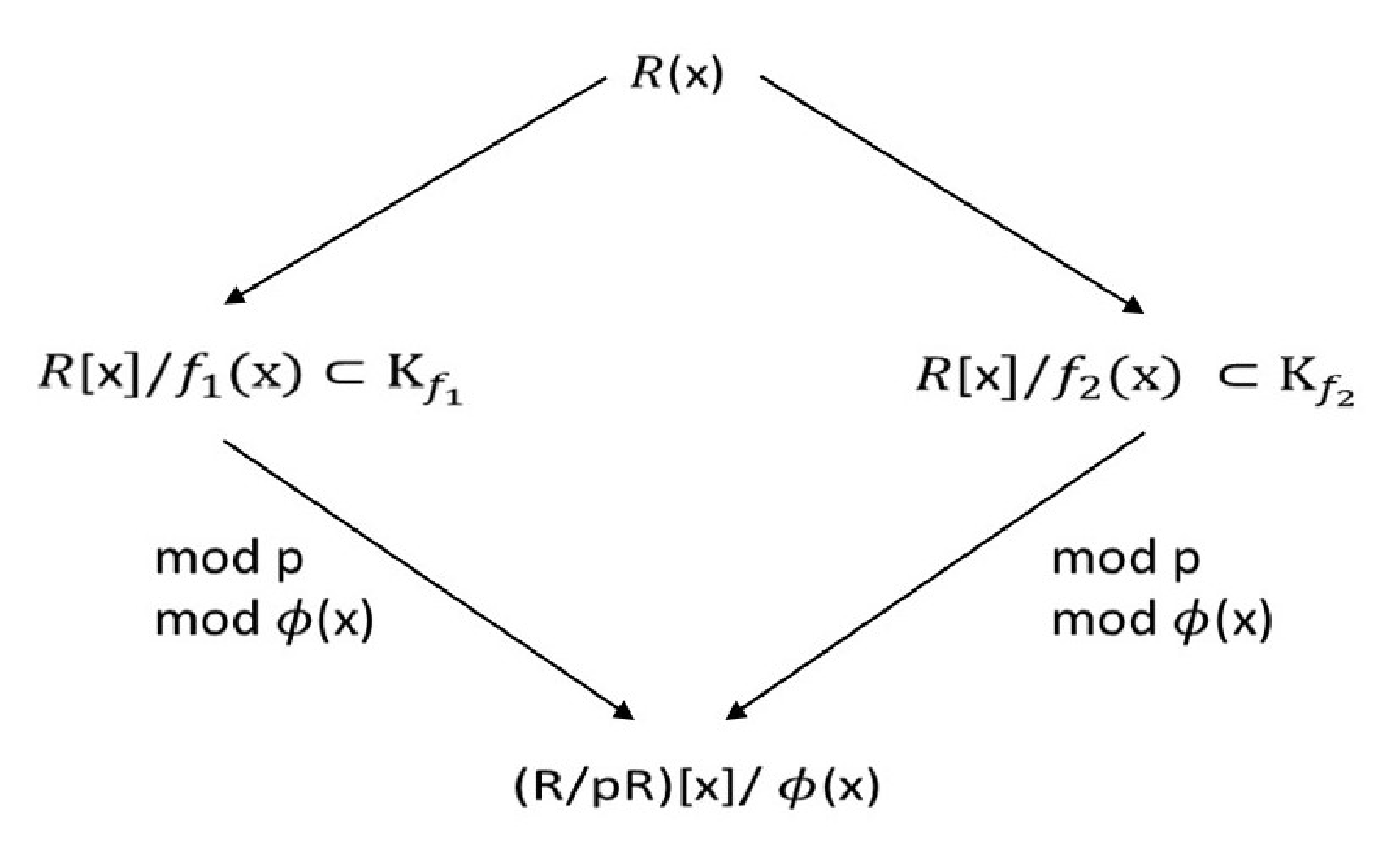}
		\caption{Commutative diagram of exTNFS. When $\phi(x) = x-z$ for some $z \in R$ this is the diagram of TNFS.}
		
	\end{figure}
	
	We explain how exTNFS is able to reduce the security over finite fields with medium characteristics as follows, we can clearly see from the conjugate method\cite{2015Improving} that the constant c in the complexity of the boundary characteristic case is lower than that of the medium characteristic case. If $F_{p^n}=F_{(p^\eta)^\kappa}$, where $p=L_Q\left(\alpha,c\right)$ for $1/3<\alpha <2/3$, then $F_{p^n}$ can be represented as $F_{q^\kappa}$, where $q=p^\eta$ and $q=L_Q\left(\alpha,c\right)$ for $\alpha=2/3$. This transforms the problem of solving the discrete logarithm of a finite field over medium characteristic into the problem of solving the discrete logarithm of a finite field over boundary characteristic, thereby reducing the complexity. The essence of this method is to reduce the size of the norm. We use $r(x)\in R(x)$ to sieve relation pairs. A relation is obtained from the polynomial $\varphi(x)$ if both the norms $N(\varphi,f_1)={Res}_t({Res}_x(\varphi(x),f_1(x)),h(t))$ and $N(\varphi,f_2)={Res}_t({Res}_x(\varphi(x),f_2(x)),h(t))$ are B-smooth. Bounds on the norm are obtained from the bounds on resultants of bivariate polynomials in $x$ and $t$.
	
	The Kim-Barbulescu\cite{2016Extended} method is restricted to composite $n$ and requires that $\eta$ and $\kappa$ are coprime. In contrast, the method proposed by Sarkar and Singh, denoted as $\mathcal{C}$\cite{2016A}, is applicable to both prime-power and non-prime-power $n$, and removes the restriction on the coefficients of $\phi(x)$ in $F_p$, thereby eliminating the condition that $\eta$ and $\kappa$ are coprime. Algorithm $\mathcal{D}$\cite{2016A2} translates Algorithm $\mathcal{A}$ to the TNFS setting, and the aforementioned polynomial selection methods are all special cases of it. The best complexity achieved for the medium prime case is $L_Q\left(1/3,\sqrt[3]{48/9}\right)$ for all composite $n$.
	
	The use of multiple number fields in conjunction with the TNFS can significantly reduce asymptotic complexity. Combining the MNFS and Algorithm $\mathcal{D}$ can further improve complexity for the medium prime case, achieving a best achievable complexity of $L_Q\left(1/3,\frac{3+\sqrt{3(11+4\sqrt6)}}{{(18(7+3\sqrt6))}^{1/3}}\right)$. The SNFS-JP\cite{2014The} algorithm demonstrates that the complexity constant $c$ for the large characteristic case is lower than that of the boundary characteristic case. By transforming the medium characteristic to the large characteristic, the SexTNFS\cite{2016Extended} algorithm can be obtained by combining the SNFS-JP and exTNFS algorithms. This algorithm uses polynomials $f_1(t, x) = P(x^\kappa + W(x) + t)$, $f_2(t, x) = x^\kappa + S(x) + t-u$, and $h$, an irreducible polynomial of degree $\eta$. Dropping $t$ is possible if $\kappa$ and $\eta$ are coprime. This approach achieves an asymptotic complexity of $L_Q\left(1/3,\sqrt[3]{32/9}\right)$ for the medium prime case. A variant of NFS which uses these polynomials has complexity at least $L_Q\left(1/3,\sqrt[3]{32/9}\right)$.

	As shown in Table \ref{tab2}, the asymptotic run times of the NFS variants for computing discrete logarithms in $F_Q$ in the medium characteristic case are $L_Q\left(1/3,c\right)$ where:
 \begin{table*}[!ht]
		\footnotesize
		\caption{Polynomial methods and constant c values used by different NFS variants.}
		\label{tab2}
		\centering
		\tabcolsep 22pt 
		\begin{tabular*}{\linewidth}{lll}
			\toprule
			Variants of the NFS & Polynomial method & Constant c \\\midrule
			NFS & Conjugation  & $\sqrt[3]{96/9}\approx2.201$\\
			Multiple NFS & MNFS and Algorithm-$\mathcal{A}$ & ${(8(9+4\sqrt6)/15)}^{1/3}\approx2.156$\\
			SNFS & SNFS-JP & $((64/9)·(\tau+1)/\tau)^{1/3}$\\
			exTNFS for some p & exTNFS and Algorithm-$\mathcal{D}$  & $\sqrt[3]{48/9}\approx1.747$\\
			multiple exTNFS for some p & Multiple exTNFS and Algorithm-$\mathcal{D}$ & $\frac{3+\sqrt{3(11+4\sqrt6)}}{{(18(7+3\sqrt6))}^{1/3}}\approx1.710$\\
			SexTNFS & exTNFS and SNFS-JP & $\sqrt[3]{32/9}\approx1.526$\\
			\bottomrule
		\end{tabular*}
	\end{table*}

\begin{figure}[htbp]
		\centering
		\includegraphics[width=0.5\textwidth]{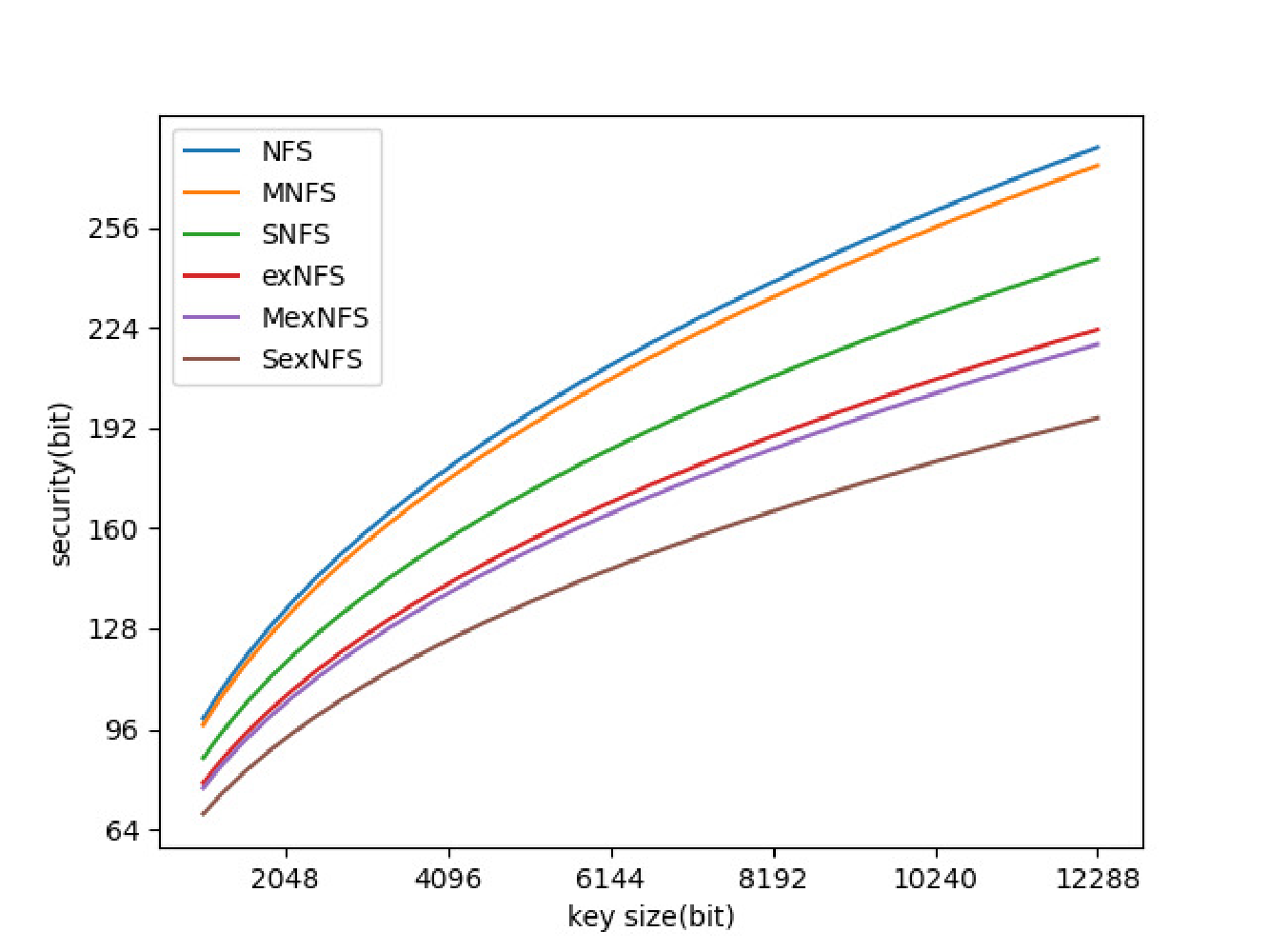}
		\caption{Comparison of the security of various NFS algorithms.}
	\end{figure}

 	\subsection{Improvement of the hidden constant o(1)}
	This subsection presents two methods, namely the Barbulescu-Duquesne method\cite{10.1007/s00145-018-9280-5} and the Guillevic-Singh method\cite{2019On}, which aim to enhance the hidden constant $o(1)$. Barbulescu and Duquesne proposed that the constant $c$ cannot be lower than 1.526, and therefore, the improvements implemented in their method are focused on improving the hidden constant $o(1)$. The hidden constant $o(1)$ is replaced by $\epsilon$ in the complexity equation $2^{\epsilon}L_Q(1/3,c)$. The estimated values of $\epsilon$ from the recently computed record are summarized in Table \ref{tab3}.

	\begin{table*}[!h]
		\footnotesize
		\caption{Value of $\epsilon$ to match the formula cost(NFS)=$2^\epsilon L_Q(1/3,c)$.}
		\label{tab3}
		\centering
		\tabcolsep 6pt 
  
		\begin{tabular*}{\linewidth}{cccccc}
			\toprule
			variant & NFS for $F_p$\cite{2020Comparing}  & composite n NFS\cite{2015Improving} & composite n MNFS\cite{2015Improving} & SNFS for $F_p$\cite{2016A12}\\\hline
			c & 1.932 & 1.747 & 1.710 & 1.526 \\
			$\epsilon$ & -10.17 & -7.000 & -7.000 & -4.500 \\
			\bottomrule
		\end{tabular*}
	\end{table*}
	
	\subsubsection{Barbulescu-Duquesne method}
	In security analysis, the norm of an algebraic number is a crucial factor. While Menezes, Sarkar and Singh\cite{2017Challenges} relied on the upper bound of the norm, Barbulescu and Duquesne observed that the majority of norms were significantly smaller than the upper bound. To address this, they utilized the average value of the norm to reduce the cost of the linear algebra stage and employed automorphisms and roots of unity, which were not used by Menezes, Sarkar, and Singh, to obtain more accurate lower bound security estimates. This method involved averaging 25600 norms to ensure a probability of error of $2^{-128}$.
	
	\begin{algorithm*}[!ht]
		\renewcommand{\algorithmicrequire}{\textbf{Input:}}
		\renewcommand{\algorithmicensure}{\textbf{Output:}}
		\caption{Barbulescu-Duquesne method}
		\label{alg1}
		\label{power1}
		\begin{algorithmic}[1] 
			\REQUIRE  Types of curves, $n$ and $u$ 
			\ENSURE $\mathbf{cost}$ 
			
			\STATE $Select\ \kappa\ and\ \eta\ = n/\kappa;$
			\STATE Select Polynomial $h$ of degree $\eta$ with small and few coefficients
			\STATE Select Polynomial $f_1$ and $f_2$ with SexTNFS\cite{2016A2} method
			\STATE cost = $2^{256}$
			\FOR{$log_{2}A=1$ to $100/\eta$}
			\FOR{$log_{2}B=1$ to $100$}
			\STATE random 25600 tuple $a_0,...a_{\eta-1},b_0,...b_{\eta-1} \in [-A,A]^{2\eta}\ (a_0 >= 0)$
			\STATE $N_{f_1}=\sum Res_t\left(Res_x\left(\sum_{i=0}^{\eta-1}{a_it^i-}x\ \sum_{i=0}^{\eta-1}{b_it^i},\ f_1\left(t,\ x\right)\right),\ h\left(t\right)\right)/25600$
			\STATE $N_{f_2}=\sum Res_t\left(Res_x\left(\sum_{i=0}^{\eta-1}{a_it^i-}x\ \sum_{i=0}^{\eta-1}{b_it^i},\ f_2\left(t,\ x\right)\right),\ h\left(t\right)\right)/25600$
			\STATE $p_{f_1}=\rho\left(\frac{\log_2{N_{f_1}}}{\log_2{B}}\right)$\qquad$p_{f_2}=\rho\left(\frac{\log_2{N_{f_2}}}{\log_2{B}}\right)$
			\STATE sievingspace = $(2 A + 1)^{2 \eta}/(2 w)$
			\STATE factorbase = $2 B/log(B)$
			\STATE relations = sievingspace$\ *\ p_{f_1}\ *\ p_{f_2}$
			\IF {$relations < factorbase$}
			\STATE sievecost = $\frac{2B}{\mathcal{A}\log{B}}{p_f}^{-1}{p_g}^{-1}$
			\STATE linearalgebra = $2^7\frac{B^2}{\mathcal{A}^2\left(\log{B}\right)^2\left(\log_2{B}\right)^2}$
			\IF {$sievecost1 + linearalgebra < cost$}
			\STATE cost = sievecost1 + linearalgebra;
			\ELSE
			\STATE continue;
			\ENDIF
			\ELSE
			\STATE continue
			\ENDIF
			\ENDFOR
			\ENDFOR
			
			\STATE \textbf{return} cost.
		\end{algorithmic}
	\end{algorithm*}

 \subsubsection{Guillevic-Singh method}
	Polynomial selection is a crucial step in NFS algorithm. Murphy proposed $\alpha$-value\cite{2011Polynomial,2015ROOT,1999Polynomial} to compare the quality of polynomials. Guillevic and Singh\cite{2019On} extended this concept to the TNFS and proposed a new generalized $\alpha$-function to evaluate the chosen polynomial. This function measures the deviation of the norm of the number field in terms of smoothness probability compared with a random integer of the same size. Overall, the $\alpha$-function plays a critical role in selecting high-quality polynomials for NFS. 
	\subsubsection{Comparison of these two methods}
	Both of these methods operate under the assumption that the number of relations is greater than the number of base factors in Equation (1). A and B are the bounds on the sieve polynomial and factor basis. These methods are designed to minimize the cost of relation collection and linear algebra in Equation (2) by carefully selecting values for A and B.

	$$\frac{\left(2A+1\right)^{2\eta}}{2w}\cdot p\left(\frac{\log_2{\left(N_{f_1}\right)}}{\log_2{B}}\right)\cdot p\left(\frac{\log_2{\left(N_{f_2}\right)}}{\log_2{B}}\right)\geq\frac{2B}{\log{\left(B\right)}}\ \ (1)$$
	$$\mathrm{cost\ }=c_{sieve}\frac{2B}{\mathcal{A}\log{B}}\rho\left(\frac{\log_2{\left(N_{f_1}\right)}}{\log_2{B}}\right)^{-1}\rho\left(\frac{\log_2{\left(N_{f_2}\right)}}{\log_2{B}}\right)^{-1}$$
 
 $$+c_{lin.alg}\frac{(2{B)}^2}{\mathcal{A}^2\left(\log{B}\right)^2\left(c_{filter}\right)^2}\ \ (2)$$
	
	where $w$ is the half of the number of roots of unity of $h$, $\mathcal{A}$ is the number of automorphisms of $h$ multiplied by the number of common number of automorphisms of $f$ and $g$. $\rho$ is Dickman’s function which is a special function used to estimate the proportion of smooth numbers up to a given bound.
	
	The two methods differ in their constant values $c_{sieve}$, $c_{filter}$ and $c_{lin.alg}$ for calculating cost. These values are shown in Table \ref{tab4}, where r is the order of the discrete logarithm group and a machine word length of 64. The values used by the models of Barbulescu-Duquesne \cite{2019Optimal} and Guillevic-Singh\cite{2020A12,2020Cocks}. The security calculated by the Barbulescu-Duquesne method is more conservative,in this work we use this method.
	
	\begin{table}[!h]
		\footnotesize
		\caption{parameters of BD and GS model}
		\label{tab4}
		\centering
		\tabcolsep 10pt 
		\begin{tabular}{cccc}
			\toprule
			Model & $c_{filter}$ & $c_{sieve}$ & $c_{lin.alg}$ \\\hline
			Barbulescu-Duquesne & $log_2B$ & 1 & 128 \\
			Guillevic-Singh & 20 & $loglogB$ & $200\lceil|r| / 64\rceil$ \\
			\bottomrule
		\end{tabular}
	\end{table}

	\section{Pairing-Friendly Curves}
 Depending on the required embedding degree, some families of PF curves have been built \cite{2010A}. Among the most popular families are BN, BLS and KSS curves, we enumerate several international standard curves and give the parameters.
	\subsection{BN Curves}
	A BN curve\cite{2005Pairing} is an elliptic curve E defined over a finite field $F_p$, with an embedding degree of 12. The curve's order $r$ and characteristic $p$ are parametrized by $p=36u^4+36u^3+24u^2+6u+1$, $r=36u^4+36u^3+18u^2+6u+1$ for some well chosen $u$ in $\mathbb Z$. It has an equation of the form $y^2=x^3+b$, where $b\in F_p^\ast$. In the next chapter, we will analyze the security of BN256, BN446 and BN638, whose parameters are shown in Tabel \ref{tab5}
	\begin{table*}[!h]
		\footnotesize
		\caption{BN curves.}
		\label{tab5}
		\centering
		\tabcolsep 18pt 
		\begin{tabular*}{\linewidth}{ccccc}
			\toprule
			& source & u & p & r \\\hline
			BN256 & SM9\cite{SM9} & 0x600000000058f98a & 256 & 256 \\
			BN446 & ISO/IEC\cite{15946-5:2022} & $2 ^ {110} + 2^{36} + 1 $ & 446 & 446 \\
			BN512 & ISO/IEC\cite{15946-5:2009} & 0x6882f5c030b0f7f010b306bb5e1bd80f & 512 & 512\\
			BN638 & FIDO Alliance\cite{FIDO} & $2^{158} - 2^{129} + 2^{128} - 2^{68} + 1$ & 638 & 638 \\
			\bottomrule
		\end{tabular*}
	\end{table*}

	\subsection{BLS Curves}
	BLS curves were introduced in \cite{2002Constructing}. For BLS12, the characteristic $p$ and order $r$ are parametrized by $p=\left(u-1\right)^2\left(u^4-u^2+1\right)/3+u$ and $r=u^4-u^2+1$. BLS24 has parameters $p=\left(u-1\right)^2\left(u^8-u^4+1\right)/3+u$ and $r=u^8-u^4+1$. In the next chapter, we will analyze the security of BLS12-381, BLS12-440, BLS12-462, BLS24-479 and BLS24-559, whose parameters are shown in Tabel \ref{tab7}

	\begin{table*}[!h]
		\footnotesize
		\caption{BLS curves.}
		\label{tab7}
		\centering
		\tabcolsep 18pt 
		\begin{tabular*}{\linewidth}{ccccc}
			\toprule
			curves & source & u & p & r \\\hline
			BLS12-381 & Zcach\cite{BLS12-381} & $ -(2^{63}\ +\ 2^{62} + 2^{60} + 2^{57} + 2^{48} +2^{16})$ & 381 & 255 \\
			BLS12-440 & ISO/IEC\cite{15946-5:2022} & $ -(2^{74}-2^{72}+2^{41}-2^{26}+2^{16})$ & 440 & 295 \\
			BLS12-462 & ISO/IEC\cite{15946-5:2022} & $ -2^{77}+2^{50}+2^{33}$ & 462 & 308 \\
			BLS24-479 & AMCL\cite{AMCL} & $2^{48}-2^{14}-2^{12}-2^4$ & 479 & 384 \\
			BLS24-559 & ISO/IEC\cite{15946-5:2022} & $-2^{56}-2^{43} + 2^9 - 2^6$ & 559 & 448 \\
			
			\bottomrule
		\end{tabular*}
	\end{table*}

	\subsection{KSS Curves}
	KSS curves are also available for different embedding degrees \cite{2008Constructing}. For KSS16, the characteristic $p$ and order $r$ are parametrized by $p=(u^{10}+2u^9+5u^8+48u^6+152u^5+240u^4+625u^2+2396u+3125)/980$, $r={(u}^8+{48u}^4+625)/61250$. KSS18 has parameters $p=(u^8+5u^7+7u^6+37u^5+188u^4+259u^3+343u^2+1763u+2401)/21$, $r={(u}^6+{37u}^3+343)/343$. In the next chapter, we will analyze the security of KSS18-508 and KSS18-676, whose parameters are shown in Tabel \ref{tab18}
	
	\begin{table*}[!h]
		\footnotesize
		\caption{KSS curves.}
		\label{tab18}
		\centering
		\tabcolsep 24pt 
		\begin{tabular*}{\linewidth}{ccccc}
			\toprule
			curves & source & u & p & r \\\hline
			KSS18-508 & RELIC\cite{10.1007/978-3-642-36334-4_11} & $ -2^{64}-2^{51}+2^{46}+2^{12}$ & 508 & 376 \\
			KSS18-676 & RELIC\cite{10.1007/978-3-642-36334-4_11} & $ -2^{85}-2^{31}-2^{26}+2^6$ & 676 & 501 \\
			\bottomrule
		\end{tabular*}
	\end{table*}

	\section{Security of pairing-friendly curves}

	In this chapter, first, we use Barbulescu-Duquesne method to analyze the security of various international standards for PF curves, compare the curve-side security and field-side security of each curve cluster. Then, we find the characteristic $p$ of each family of curves, under the precondition of 128bit, 160bit, 192bit and 256bit security level. Finally, the efficiency of pairing is calculated through these characteristic, and the most appropriate curve under each security level is obtained.
	
	\subsection{Security of BN curves}
	BN256 is used in SM9\cite{SM9}. First we choose parameters and polynomials, $\kappa=2$ and $\eta=6$. $h=t^6-t^3-t-1$, $f_1=P\left(x^2+t\right)$, $f_2=x^2+t-u$, $P\left(u\right)=36u^4+36u^3+24u^2+6u+1$, $w=1$, $\mathcal{A}=2$. Then, the bound A on the coefficients of the sieve polynomial and the bound B on the factor basis can be obtained by Algorithm \ref{power1}, $A=176$, $log_2B=57.6$. sieving space is$\mathrm{\ }\left(2A+1\right)^{2\eta}/\left(2w\right)=2^{100.57}$, the norm of $f_1$ and $f_2$ are $\log_2{\left(N_{f_1}\right)}\approx424.8$, $\log_2{\left(N_{f_2}\right)}\approx466.51$.  The probability of being smooth is given by $\rho\left(\frac{\log_2{\left(N_{f_1}\right)}}{\log_2{\left(B\right)}}\right)\approx2^{-21.88}$, $\rho\left(\frac{\log_2{\left(N_{f_2}\right)}}{\log_2{\left(B\right)}}\right)\approx2^{-25.37}$, 
	Relation is $2^{100.57-21.88-25.37}\approx2^{53.32}$. Factor basis is given by $2B/\log{\left(B\right)}\approx2^{53.28}$. Finally, the formula (1) is satisfied, we can calculate the result using the formula (2). Adding the cost of relations to the cost of linear algebra gives a total cost of $2^{99.92}$. BN256 offers 100 bits of security.

	\textbf{BN446 in ISO/IEC\cite{15946-5:2022}}: the parameters and polynomials are all the same as those of BN256. We can get $A=993$, $log_2B=72.2$, $\log_2{\left(N_{f_1}\right)}\approx538.37$, $\log_2{\left(N_{f_2}\right)}\approx799.88$. The total cost is $2^{128.45}$. BN446 offers 129 bits of security. 
	
	\textbf{BN512 in ISO/IEC\cite{15946-5:2009}}: We can get $A=1541$, $log_2B=76.6$, $\log_2{\left(N_{f_1}\right)}\approx572.7$, $\log_2{\left(N_{f_2}\right)}\approx888.52$. The total cost is $2^{137.19}$. 
	BN512 offers 138 bits of security. 
	
	\textbf{BN638 in FIDO Alliance\cite{FIDO}}: The parameters and polynomials are all the same as those of BN256. We can get $A=3666$, $log_2B=85.2$, $\log_2{\left(N_{f_1}\right)}\approx633.42$, $\log_2{\left(N_{f_2}\right)}\approx1091.2$. The total cost is $2^{152.57}$. BN638 offers 153 bits of security. By analyzing the curve side and field side of the BN curve.
	
	As shown in Figure 5, we find that the security strength of the curve side is always larger than the security strength of the field side, so the security of the BN curve can be obtained by computing the security of the field side.
	
	\begin{figure}[htbp]
		\centering		\includegraphics[width=0.52\textwidth]{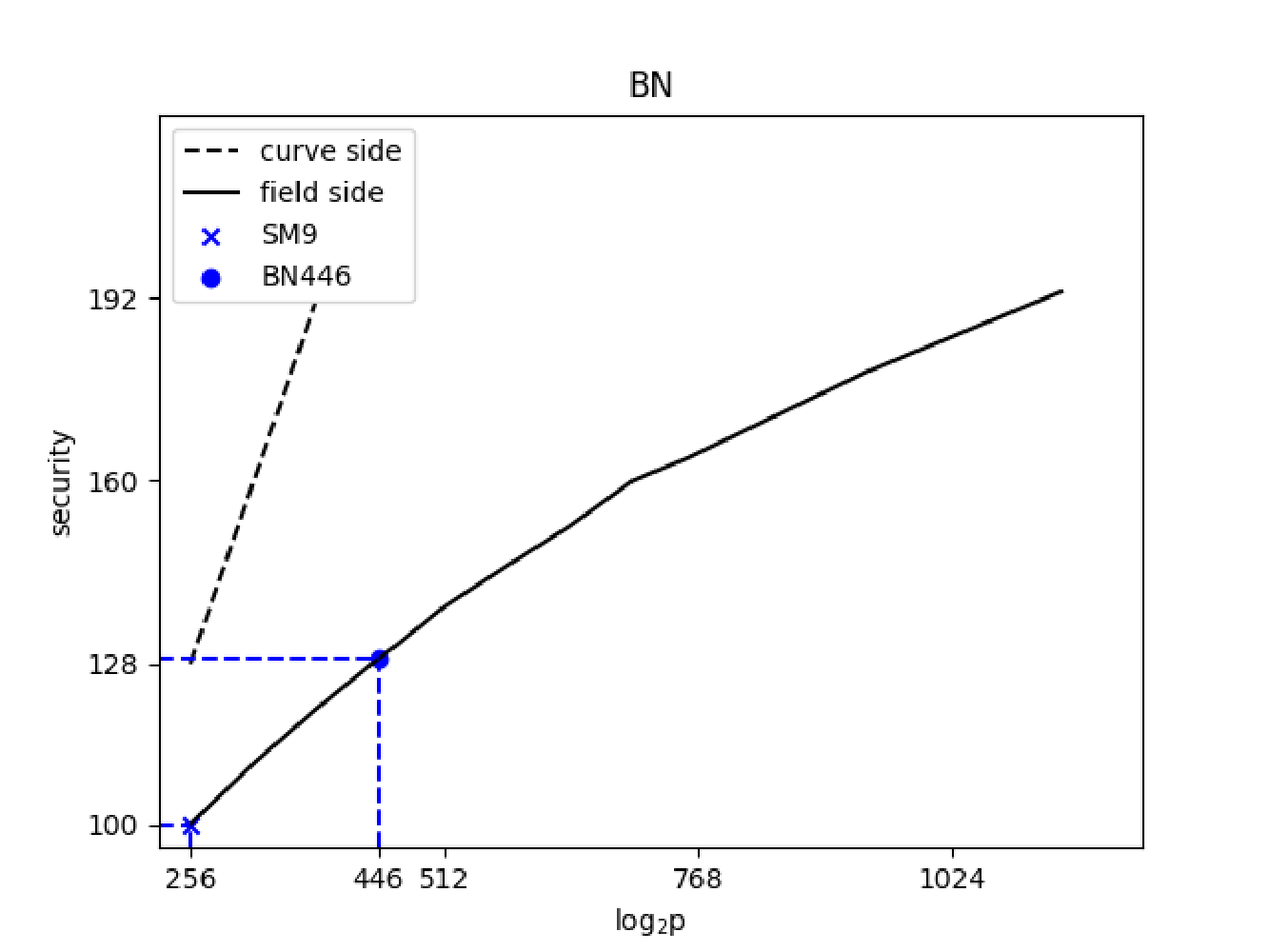}
		\label{fig50}
		\caption{Security on the curve side and field side of BN curve.}
	\end{figure}

                \begin{figure*}[t]
				\begin{minipage}{0.48\linewidth}
					\centering
					\vspace{3pt}  
					\centerline{\includegraphics[width=1.1\textwidth]{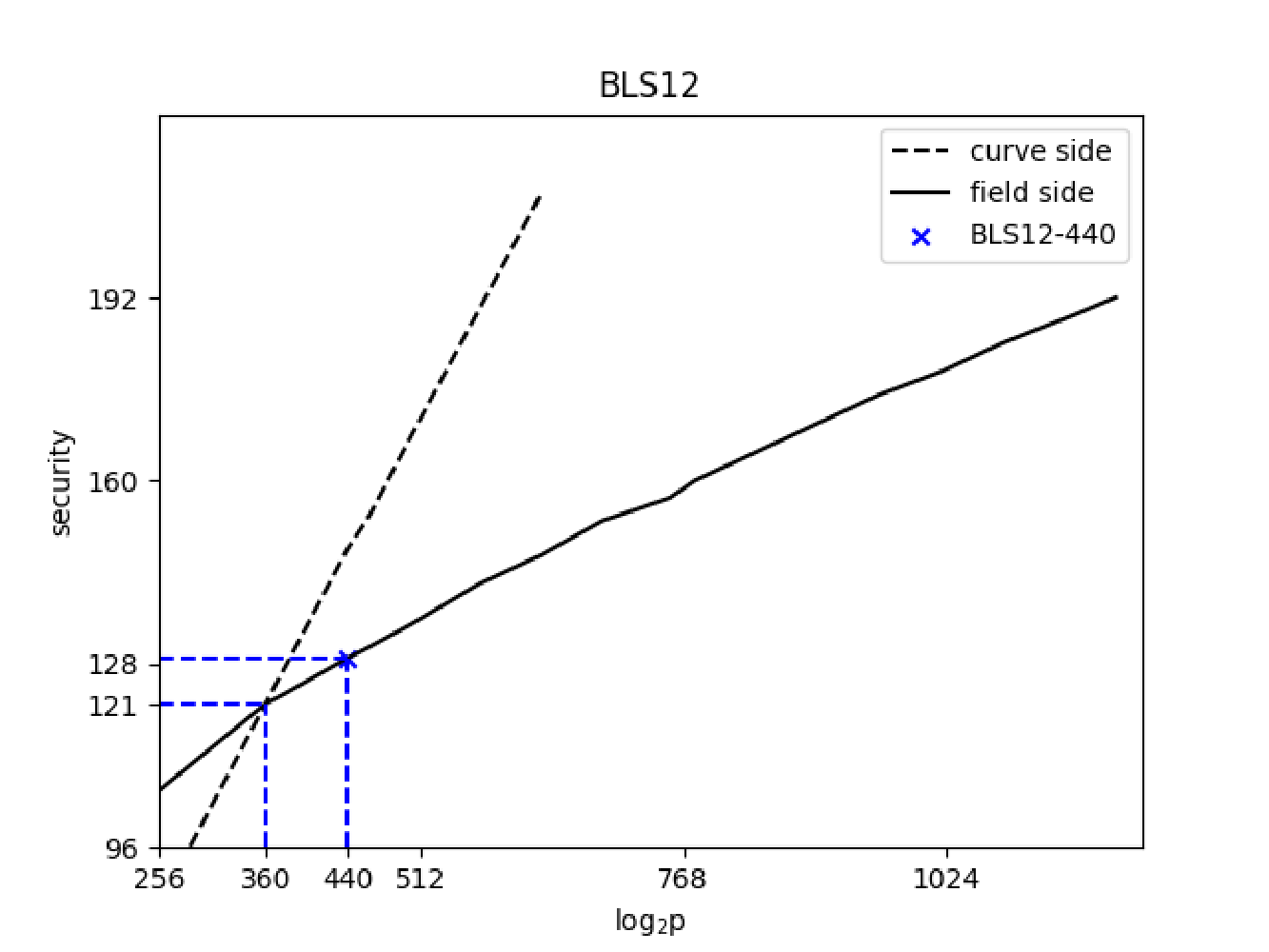}}
					\vspace{3pt}
					\centerline{\includegraphics[width=1.1\textwidth]{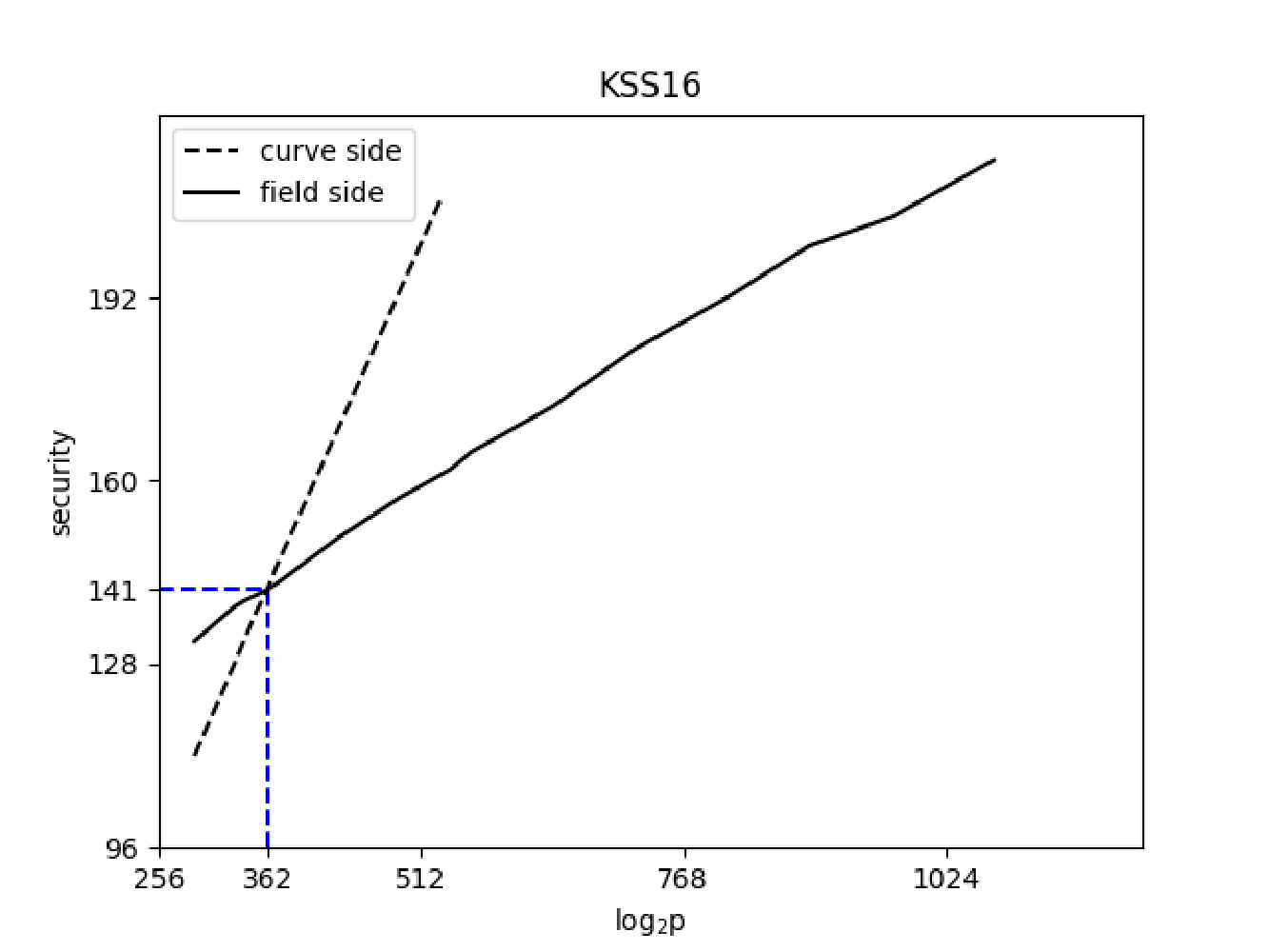}}
				\end{minipage}
					\begin{minipage}{0.48\linewidth}
						\vspace{3pt}
						\centerline{\includegraphics[width=1.1\textwidth]{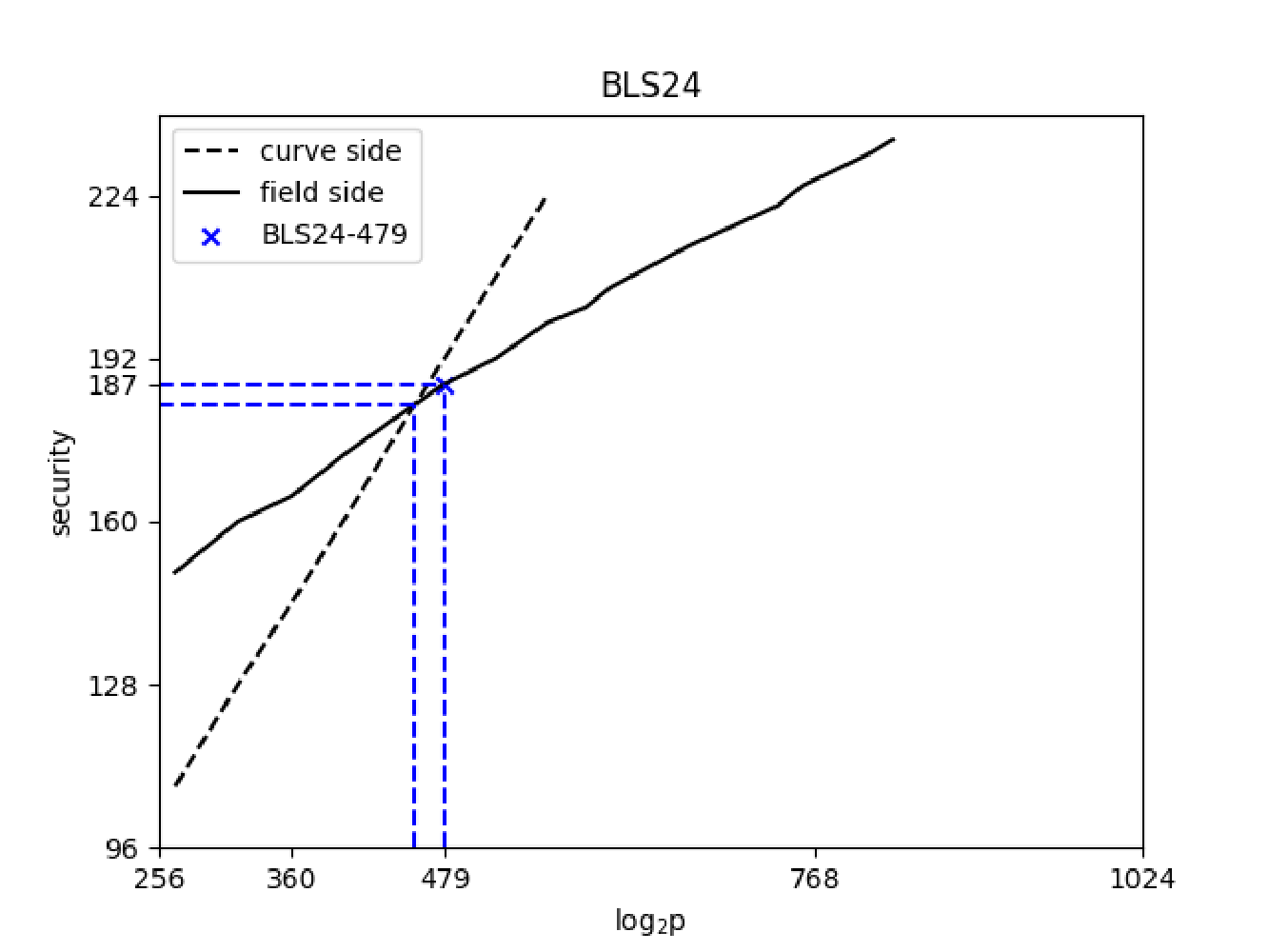}}
						\vspace{3pt}
						\centerline{\includegraphics[width=1.1\textwidth]{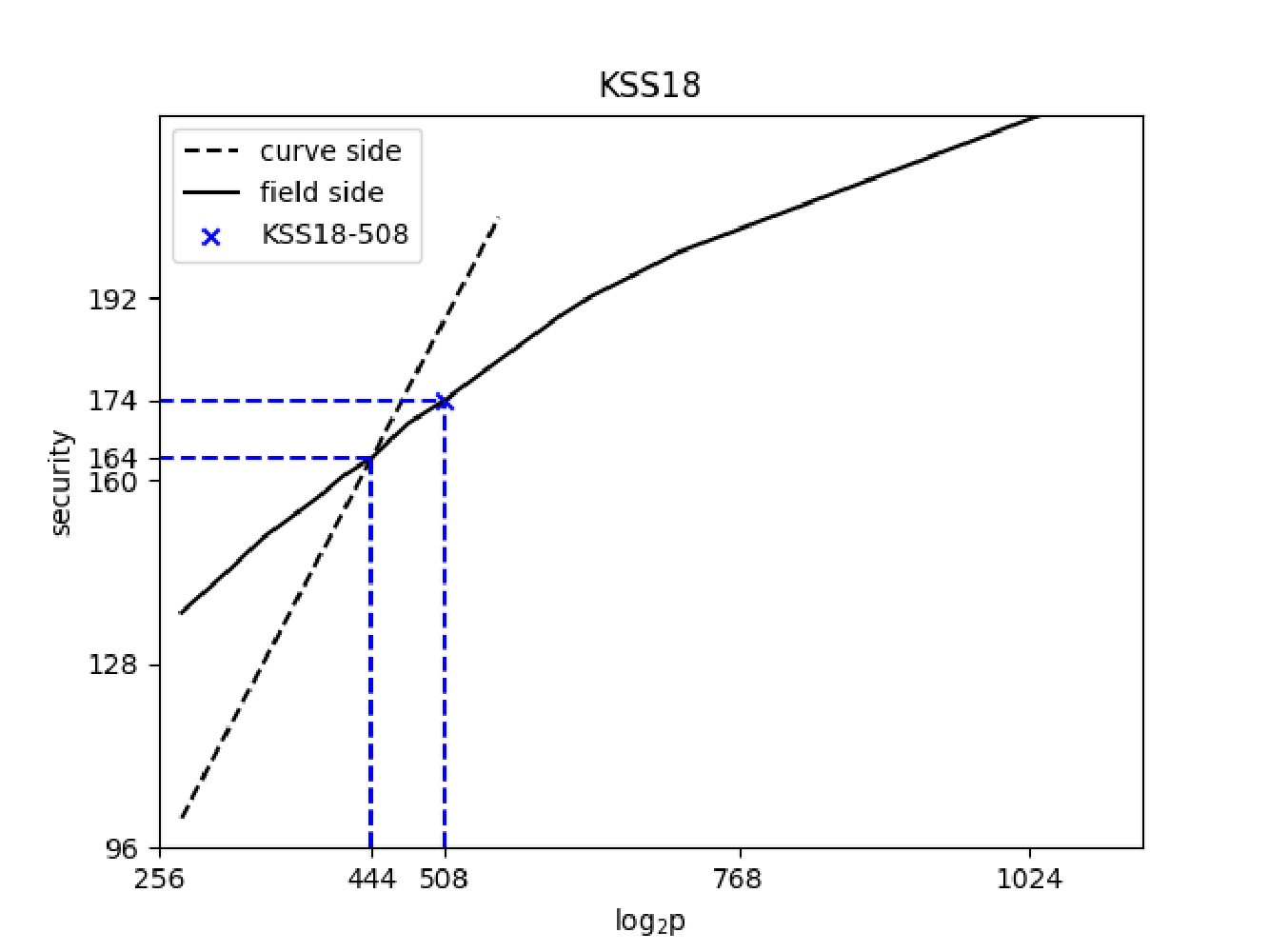}}
					\end{minipage}
					\caption{Security on the curve side and field side of BLS and KSS.}
					\label{fig40}
				\end{figure*}

     \begin{figure*}[htbp]
			\begin{minipage}{0.48\linewidth}
				\centering
				\vspace{3pt}  
				\centerline{\includegraphics[width=1.1\textwidth]{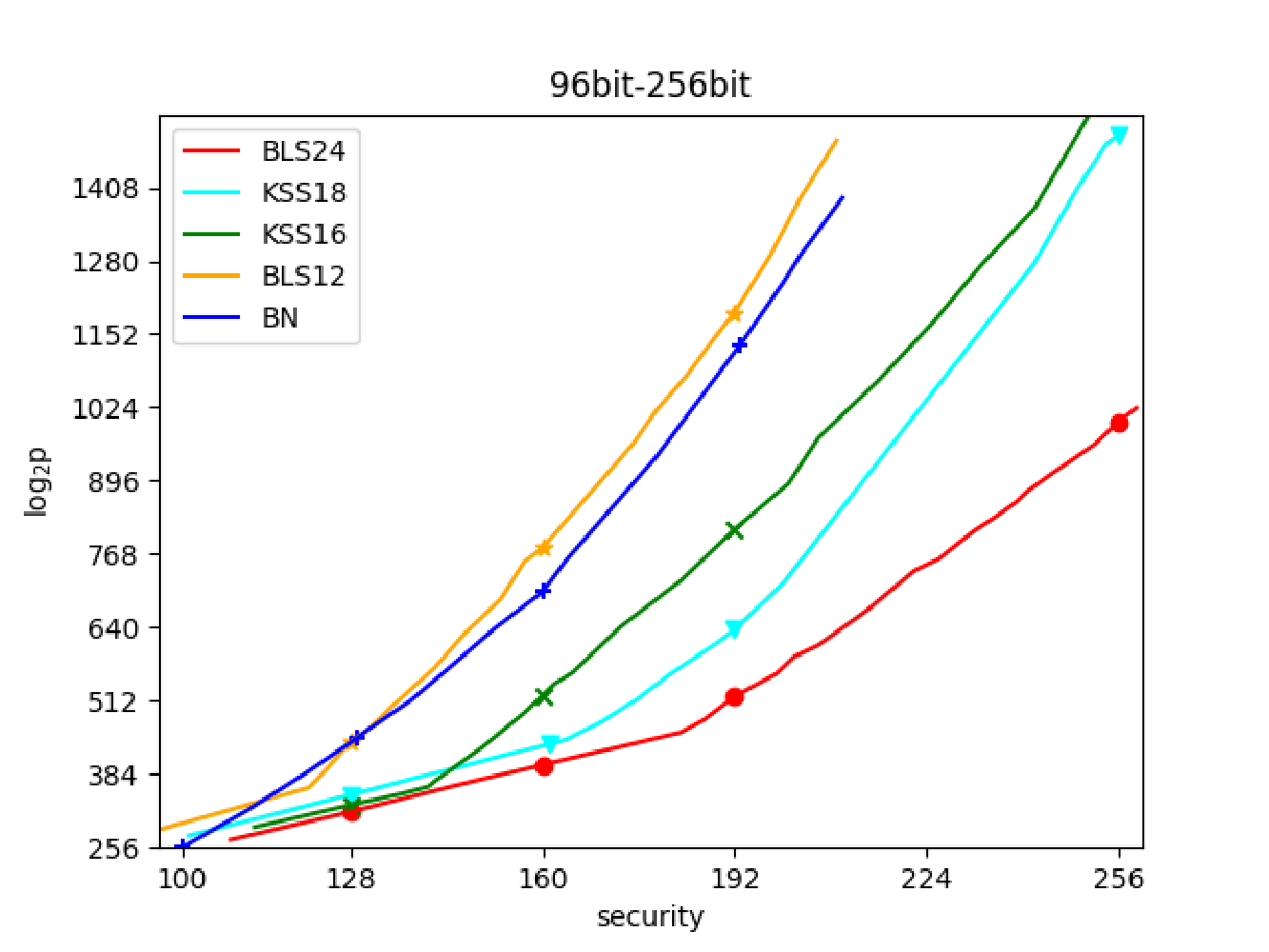}}
			\end{minipage}
				\begin{minipage}{0.48\linewidth}
					\vspace{3pt}
					\centerline{\includegraphics[width=1.1\textwidth]{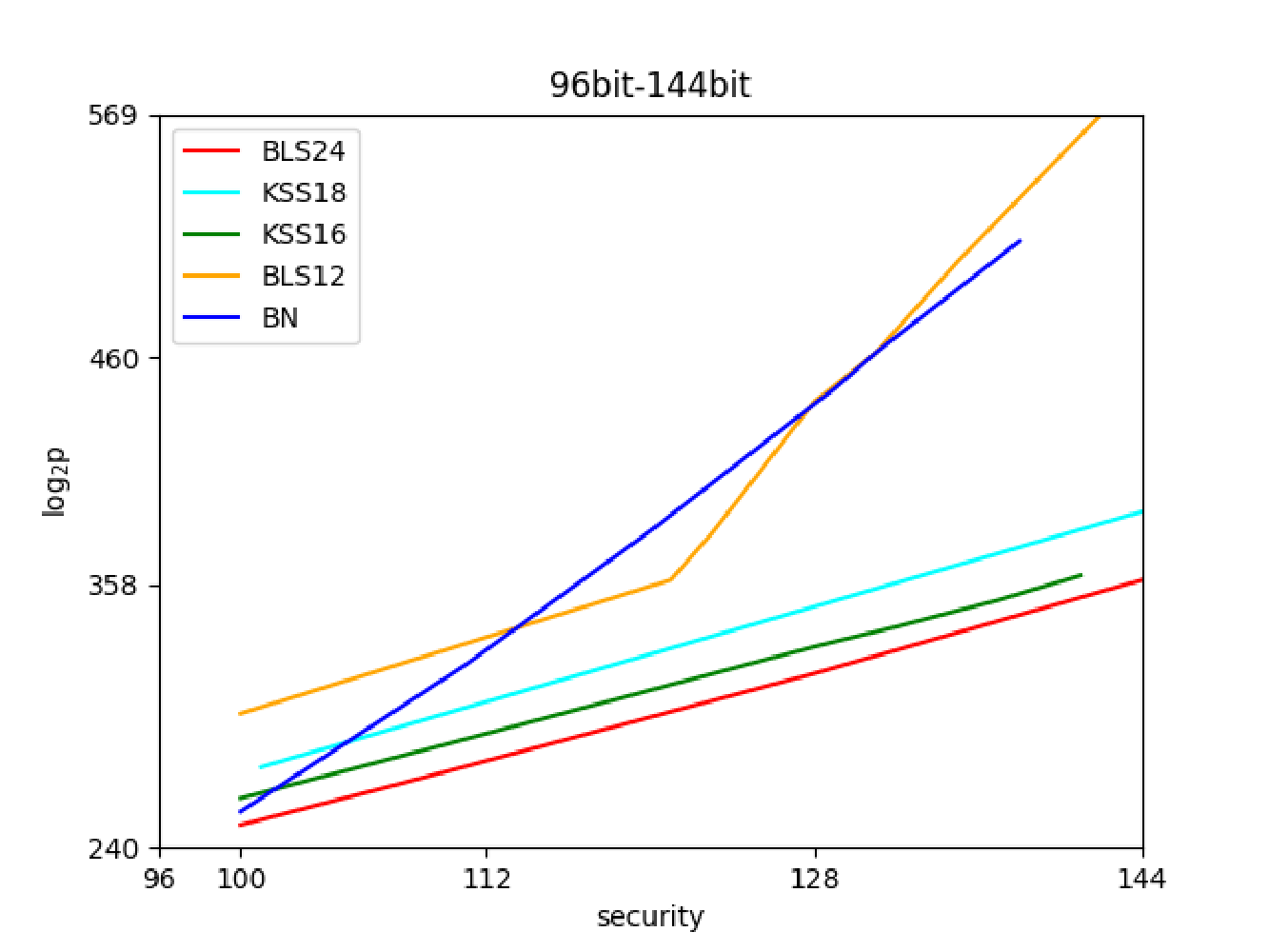}}
				\end{minipage}
				\caption{Comparison of p under specific security for each curve.}
				\label{fig8}
			\end{figure*}
   
        \subsection{Security of BLS curves}       
	BLS12-381\cite{BLS12-381} is an elliptic curve designed by Sean Bowe in 2017 to update the Zcash protocol, which can be used to efficiently build zkSnarks. First we choose parameters and polynomials, $\kappa=2$ and $\eta=6$. $h=t^6-t-1$, $f_1=P\left(x^2+t+t^2+t^4+1\right)$, ${f_2}=x^2+t+t^2+t^4+1-u$, $P\left(u\right)=\left(u-1\right)^2\left(u^4-u^2+1\right)+3u$, $w=1$, $\mathcal{A}=2$. Then, $A$, $B$ can be computed according to Algorithm \ref{power1}. $A=639$, $log_2B=68.9$. sieving space is$\mathrm{\ }\left(2A+1\right)^{2\eta}/\left(2w\right)=2^{122.85}$, 
	the norm of $f_1$ and $f_2$ are $\log_2{\left(N_{f_1}\right)}\approx725.19$, $\log_2{\left(N_{f_2}\right)}\approx494.18$. The probability of being smooth is given by $\rho\left(\frac{\log_2{\left(N_{f_1}\right)}}{\log_2{\left(B\right)}}\right)\approx2^{-37.15}$, $\rho\left(\frac{\log_2{\left(N_{f_2}\right)}}{\log_2{\left(B\right)}}\right)\approx2^{-20.49}$ ,
	Relation is $2^{122.85-37.15-50.49}\approx2^{65.21}$. Factor basis is given by $2B/\log{\left(B\right)}\approx2^{65.20}$. The formula (1) is satisfied, we can calculate the result using the formula (2). Adding the cost of relations to the cost of linear algebra gives a total cost of $2^{122.54}$. The security of BLS12-381 is 123bit.
	
	\textbf{BLS12-440 in \cite{15946-5:2022}}: The parameters and polynomials are all the same as those of BLS12-381. We can get $A=929$, $log_2B=72.8$, $\log_2{\left(N_{f_1}\right)}\approx765.72$, $\log_2{\left(N_{f_2}\right)}\approx559.31$, The total cost is $2^{128.79}$. The security of BLS12-440 is 129bit. 
	
	\textbf{BLS12-462 in \cite{15946-5:2022}}: the parameters and polynomials are all the same as those of BLS12-381. We can get $A=1067$, $log_2B=73.6$, $\log_2{\left(N_{f_1}\right)}\approx780.93$, $\log_2{\left(N_{f_2}\right)}\approx582.23$, The total cost is $2^{130.98}$. The security of BLS12-462 is 131bit. 
	
	\textbf{BLS24-479 in Apache Milagro Crypto Library\cite{AMCL}}: First we choose parameters and polynomials, $\kappa=1$ and $\eta=24$. $h=t^{24}+t^4-t^3-t-1$, $f_1=P\left(x\right)$, $f_2=x-u$, $P\left(x\right)=\left(x-1\right)^2\left(x^8-x^4+1\right)+3x$, $w=1$, $\mathcal{A}=1$. Then, we can get $A=7$, $log_2B=103$, $\log_2{\left(N_{f_1}\right)}\approx1155.6$, $\log_2{\left(N_{f_2}\right)}\approx1254.6$,  The total cost is $2^{186.25}$. The security of BLS24-479 is 187bit. 
	
	\textbf{BLS24-559 in \cite{15946-5:2022}}: The parameters and polynomials are all the same as those of BLS24-479. We can get $A=9$, $log_2B=108.6$, $\log_2{\left(N_{f_1}\right)}\approx1248.04$, $\log_2{\left(N_{f_2}\right)}\approx1455.46$, The total cost is $2^{200.82}$. The security of BLS24-559 is 201bit.
	
	As shown in Figure \ref{fig40}, by analyzing the curve side and field side of the BLS curve, when the characteristic p of BLS12, BLS24 curves is 360bit, 456bit, their security on the curve side is equal to the security on the field side. BLS12 is best suited for security strength of 121bit and BLS24 is best suited for security strength of 183bit.

		\subsection{Security of KSS curves}
		\textbf{KSS18-508 in RELIC Library\cite{10.1007/978-3-642-36334-4_11}}: First we choose parameters and polynomials, $\kappa=1$ and $\eta=18$. $h=t^18-t^4-t^2-t-1$, ${f_1}=x^8+5x^7+7x^6+37x^5+188x^4+259x^3+343x^2+1763x+2401$, ${f_2}=x-u-2$, $w=1$, $\mathcal{A}=1$. Then, $A$, $B$ can be computed according to Algorithm 1. $A=14$, $log_2B=98.4$. sieving space is$\mathrm{\ }\left(2A+1\right)^{2\eta}/\left(2w\right)=2^{173.89}$, the norm of $f_1$ and $f_2$ are $\log_2{\left(N_{f_1}\right)}\approx883.51$, $\log_2{\left(N_{f_2}\right)}\approx1240.57$.  The probability of being smooth is given by $\rho\left(\frac{\log_2{\left(N_{f_1}\right)}}{\log_2{\left(B\right)}}\right)\approx2^{-31.51}$, $\rho\left(\frac{\log_2{\left(N_{f_2}\right)}}{\log_2{\left(B\right)}}\right)\approx2^{-52.20}$ ,
		Relation is $2^{173.89-31.51-52.20}\approx2^{90.17}$. Factor basis is given by $2B/\log{\left(B\right)}\approx2^{89.76}$. The formula (1) is satisfied, we can calculate the result using the formula (2). Adding the cost of relations to the cost of linear algebra gives a total cost of $2^{173.78}$. The security of KSS18-508 is 174bit.

		\textbf{KSS18-676 in RELIC Library\cite{10.1007/978-3-642-36334-4_11}}: The parameters and polynomials are all the same as those of KSS18-508. We can get $A=23$, $log_2B=107$, $\log_2{\left(N_{f_1}\right)}\approx980$, $\log_2{\left(N_{f_2}\right)}\approx1629$. The total cost is $2^{197.69}$. The security of KSS18-676 is 198bit. 
		
		As shown in Figure \ref{fig40}, when the characteristic p of KSS16, KSS18 curves is 362bit, 444bit their security on the curve side is equal to the security on the field side. KSS16 is best suited for security strength of 141bit and KSS18 is best suited for security strength of 164bit.

\begin{figure*}[t]
				\begin{minipage}{0.48\linewidth}
					\centering
					\vspace{3pt}  
					\centerline{\includegraphics[width=1.1\textwidth]{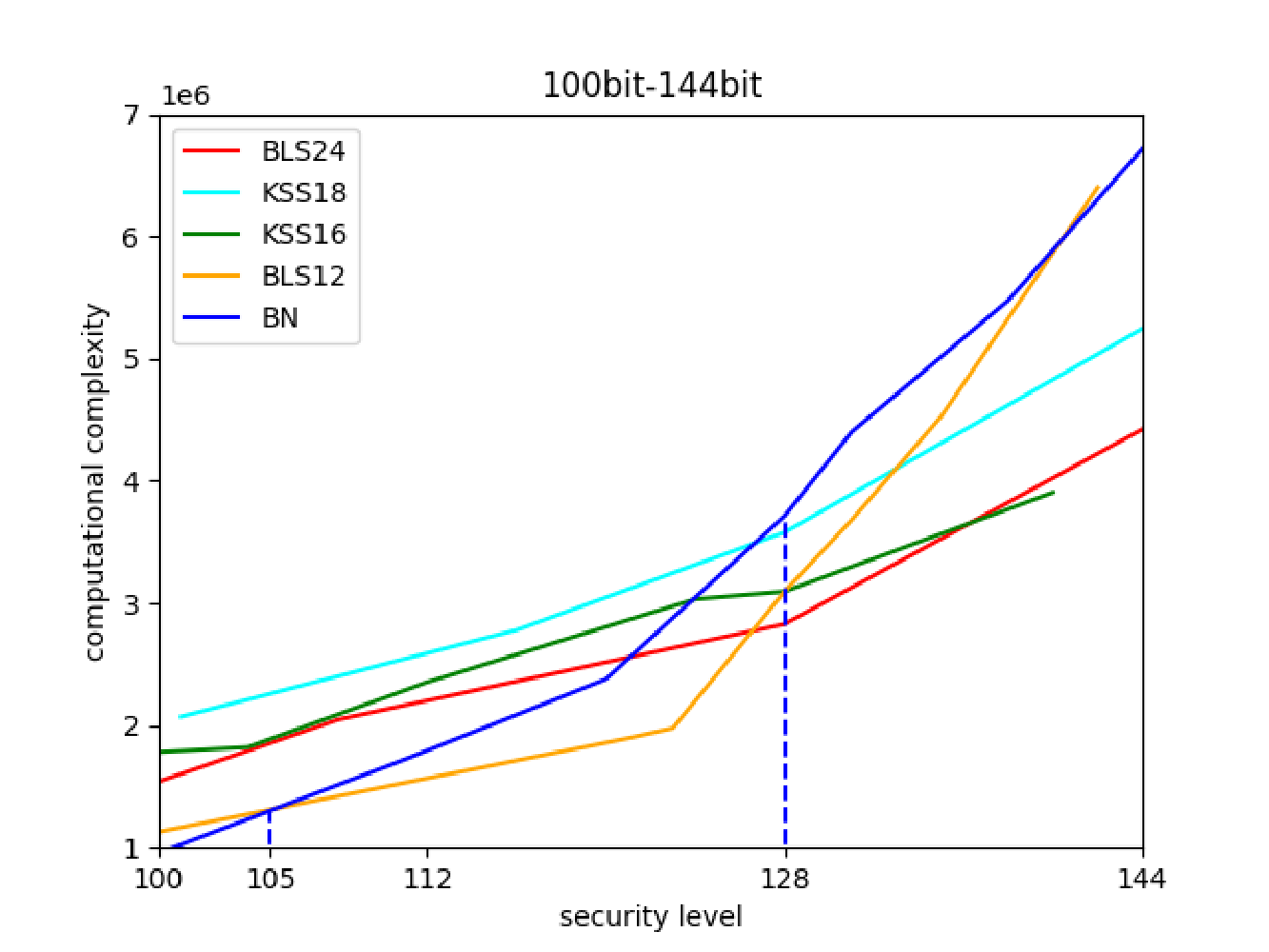}}
					\vspace{3pt}
					\centerline{\includegraphics[width=1.1\textwidth]{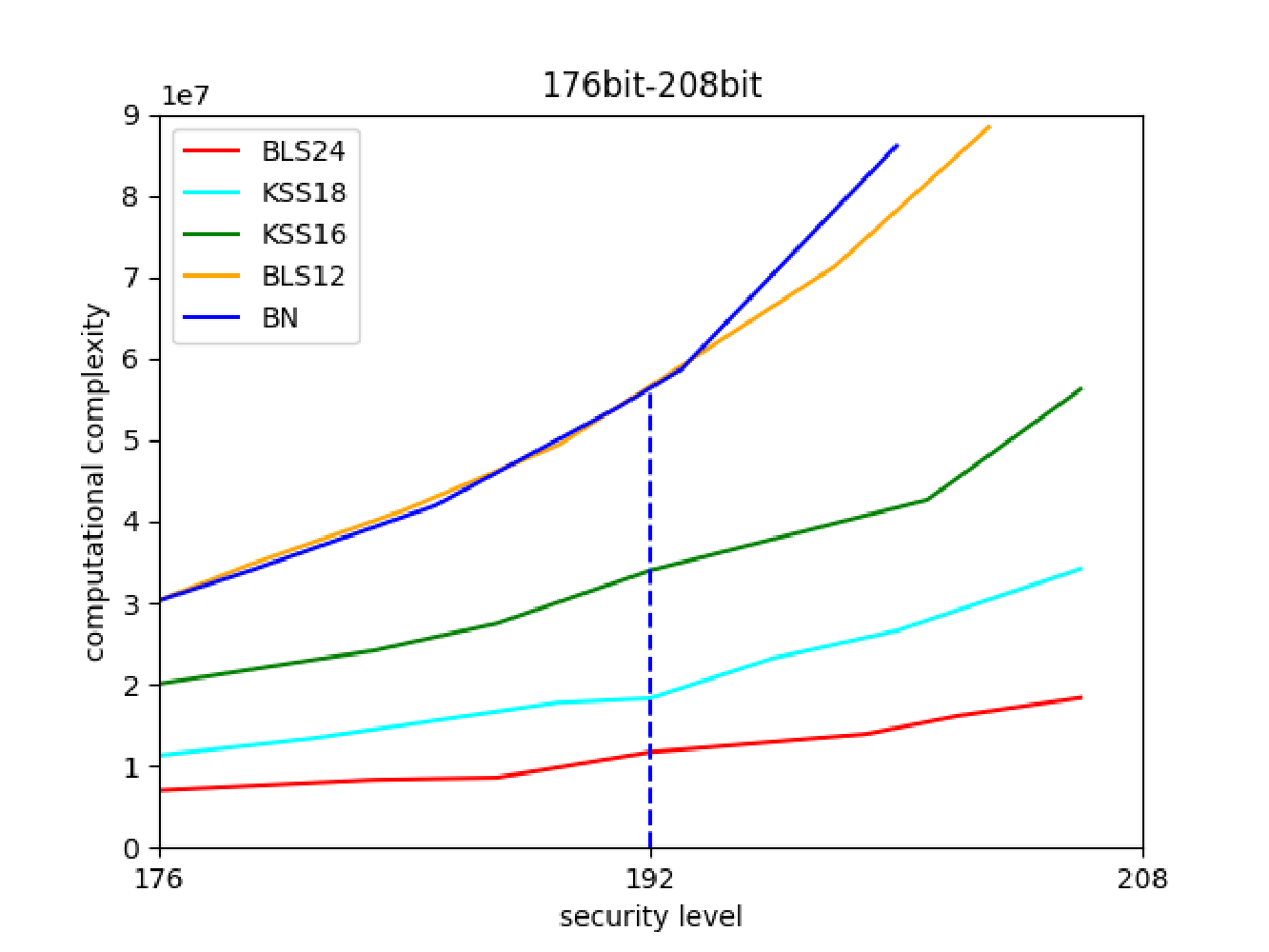}}
				\end{minipage}
					\begin{minipage}{0.48\linewidth}
						\vspace{3pt}
						\centerline{\includegraphics[width=1.1\textwidth]{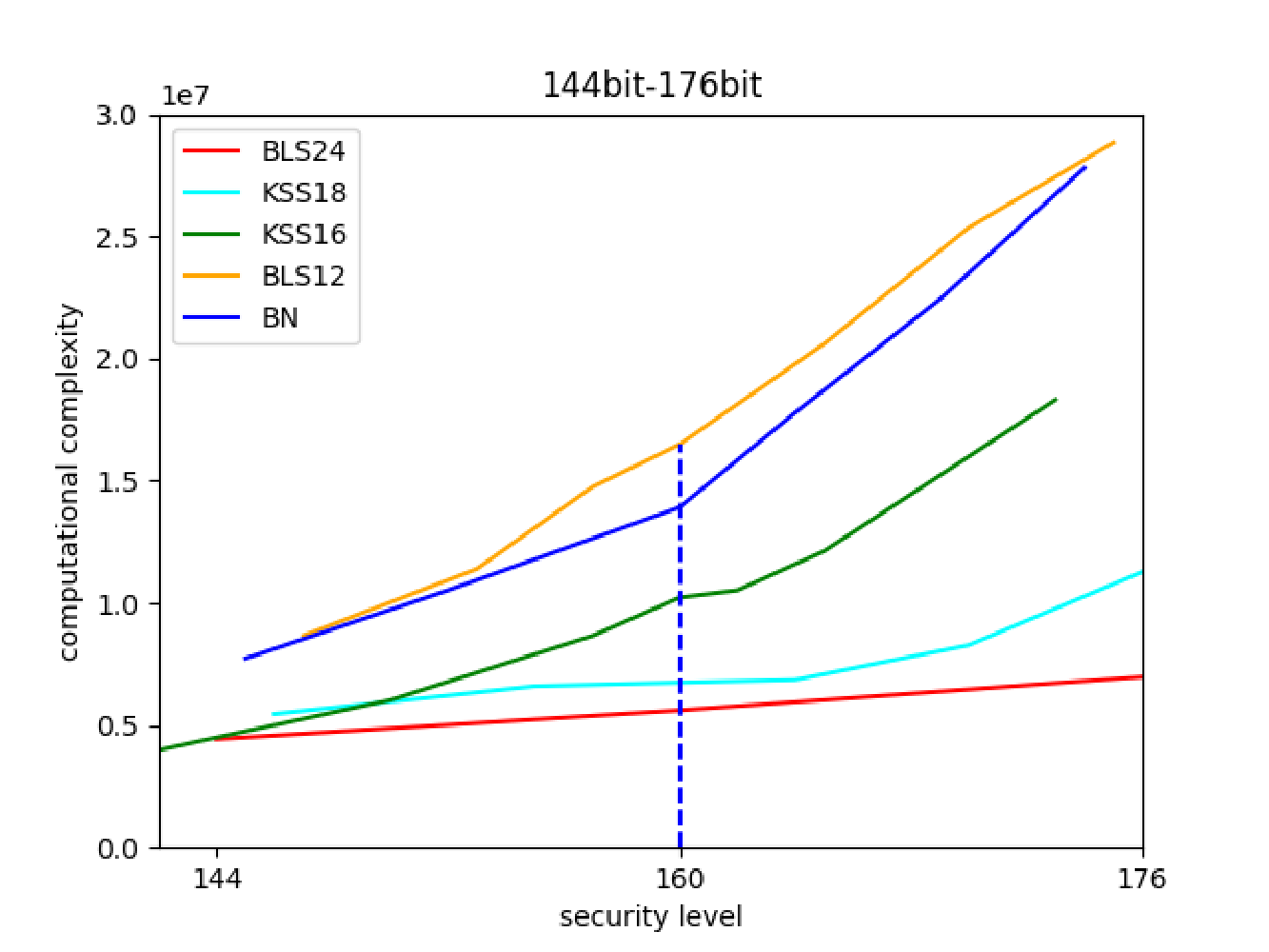}}
						\vspace{3pt}
						\centerline{\includegraphics[width=1.1\textwidth]{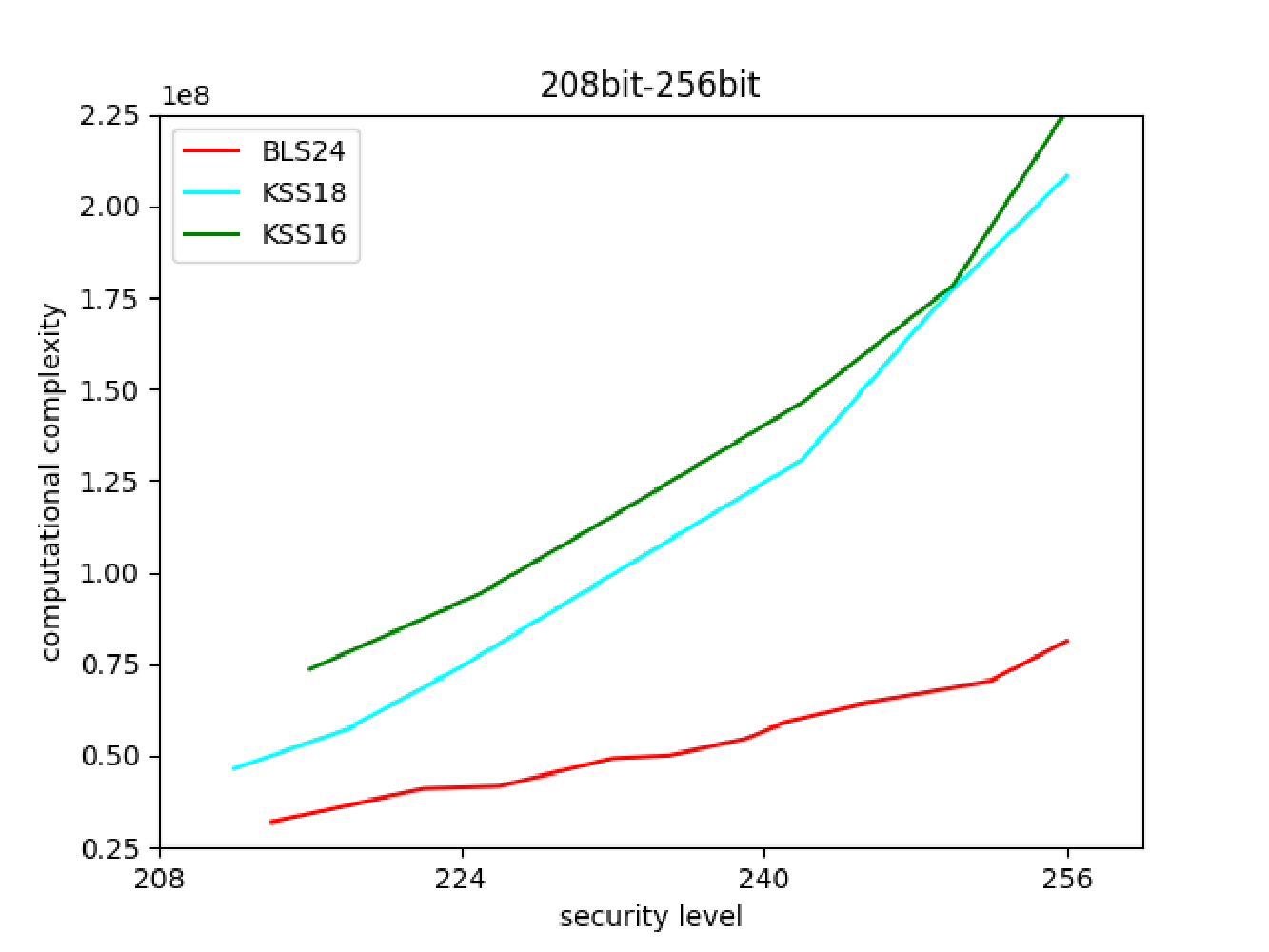}}
					\end{minipage}
					\caption{Computational complexity of different curves at specific security strength.}
					\label{Fig9}
				\end{figure*}
   
			\begin{table*}[!h]
				\footnotesize
				\caption{Security of popular pairing-friendly curves.}
				\label{tab67}
				\centering
				\tabcolsep 22pt 
				\begin{tabular*}{\linewidth}{cccccc}
					\toprule
					curve & A & $log_2(B)$ & $log_2(N_{f_1})$ & $log_2(N_{f_2})$ & security\\\hline
					BN256 & 176 & 57.6 & 424.80 & 466.51 & 100\\
					BN446 & 993 & 72.2 & 538.37 & 799.88 & 129\\
					BN512 & 1541 & 76.6 & 572.71 & 888.52 & 138\\
					BN638& 3666 & 85.2 & 633.42 & 1091.20 & 153\\
					BLS12-381 & 639 & 68.9 & 725.19 & 494.18 & 123\\
					BLS12-440 & 929 & 72.8 & 765.72 & 559.31 & 129\\
					BLS12-462 & 1067 & 73.6 & 780.93 & 582.23 & 131\\
					BLS24-479 & 7 & 103.0 & 1155.60 & 1254.6 & 187\\
					BLS24-559 & 9 & 108.6 & 1248.04 & 1455.46 & 187\\
					KSS18-508 & 14 & 98.4 & 883.51 & 1240.57 & 174\\
					KSS18-676 & 23 & 107.0 & 980.63 & 1627.97 & 198\\
					\bottomrule
				\end{tabular*}
			\end{table*}
			\subsection{Comparison of security and efficiency}
			In this subsection, we calculate and compare the size of the characteristics of each curve at the same security level, by calculating the efficiency of each curve under a specific security level, we get the most appropriate curve when the security level is 128bit, 160bit, 192bit and 256bit.


			As shown in Figure \ref{fig8}, the BN, BLS12, and KSS16 curves require a minimum characteristic p of 446, 440, and 330 bits, respectively, to attain a 128-bit security level. Although classical NFS algorithm can provide 128-bit security for both curve and field sides of the BN curve with a 256-bit p, the Barbulescu-Duquesne method demonstrates that the field side's security does not reach 100 bits. Consequently, the SM9 does not meet the 128-bit security standard.

				\begin{table}[!h]
					\footnotesize
					\caption{The characteristics(bit) of each curve at 128-bit, 160-bit, 192-bit and 256-bit security level.}
					\label{tab66}
					\centering
					\tabcolsep 10pt 
					\begin{tabular}{ccccc}
						\toprule
						curve & 128-bit & 160-bit & 192-bit & 256-bit\\\hline
						BN & 446 & 702 & 1134 & - \\ 
						BLS12 & 440 & 779 & 1190 & - \\
						KSS16 & 330 & 521 & 811 & - \\
						KSS18 & 348 & 436 & 636 & 1500\\
						BLS24 & 318 & 400 & 519 & 999\\
						\bottomrule
					\end{tabular}
				\end{table}

KSS18 and BLS24 with 160bit and 192bit security levels, exhibit a smaller characteristic p compared to other curves, indicating greater efficiency. The KSS18 curve requires p to be 434bit and 636bit to achieve 160bit and 192bit security, respectively, while the BLS24 curve requires p to be 400bit and 519bit for the same security levels. Notably, the BLS24 curve is the most suitable for 256bit security, with p requiring 999 bits. These are summarized in table \ref{tab66}

We evaluated the efficiency of pairings using the method outlined in \cite{10.1007/s00145-018-9280-5}. Our analysis involved computing the miller loop and final exponentiation for each PF curve under 100, 128, 160, 192, and 256bit security levels. Figure \ref{Fig9} indicate that the BN curve is the most efficient at 100-bit security level, followed by BLS12. At 128bit security level, BLS24 is the most efficient, followed by BLS12 and KSS16. ISO/IEC BN512 curve can achieve a security strength of 138 bits, its speed and efficiency are relatively low. At 160bit and 192bit security levels, BLS24 outperforms BN and BLS12 curves, which exhibit significant reductions in efficiency. Finally, under 256-bit security level, BLS24 is the most efficient curve, with other curves lagging far behind.
				
\section{Conclusion}
This paper provides an overview of the NFS algorithms for the PBC security evolution from three aspects:  the degree $\alpha$, the constant $c$, and the hidden constant o(1). Our review indicates that the special extended tower NFS algorithms can give the minimal constant $c$, while further research is required to optimize the hidden constant. We also evaluate the security of PF curves using the latest algorithm and find that the SM9 curve falls short of the 128-bit security threshold. Additionally, we analyze the curve-side and field-side security of BN, BLS, and KSS curves, and determine the minimum characteristics required for each curve to achieve 128, 160, 192, and 256-bit security levels. We observe that the BN curve exhibits the highest efficiency for security requirements below 105 bits, while the BLS12 curve is the most efficient choices for security levels between 105 and 128 bits. For security requirements exceeding 128 bits, the BLS24 curve is the most efficient choices.

\bibliographystyle{alpha}
\bibliography{SAPBC} 

\end{document}